\documentclass[prb,twocolumn,english,superscriptaddress,amsmath,amssymb,floatfix]{revtex4-1}

\usepackage{float}
\usepackage{graphicx}
\usepackage{dcolumn}
\usepackage{bm}

\usepackage{color}
\usepackage[colorlinks,bookmarks=false,citecolor=darkblue,linkcolor=darkblue,urlcolor=blue]{hyperref}

\definecolor{darkred}{rgb}{0.7,0.0,0.0}

\definecolor{darkblue}{rgb}{0,0.02,0.45}

\definecolor{darkgreen}{rgb}{0.02,0.45,0.0}

\definecolor{violet}{rgb}{0.8,0.2,0.6}

\graphicspath{{graphs/}}

\begin{document}

\title{Metamagnetism in the high-pressure tetragonal phase of UTe$_2$}

\author{T. Thebault}
 \affiliation{Laboratoire National des Champs Magn\'{e}tiques Intenses - EMFL, CNRS, Univ. Grenoble Alpes, INSA-T, Univ. Toulouse 3, 31400 Toulouse, France}
\author{D. Braithwaite}
 \affiliation{Univ. Grenoble Alpes, CEA, Grenoble INP, IRIG, PHELIQS, 38000, Grenoble, France}
\author{G. Lapertot}
 \affiliation{Univ. Grenoble Alpes, CEA, Grenoble INP, IRIG, PHELIQS, 38000, Grenoble, France}
\author{D. Aoki}
 \affiliation{Institute for Materials Research, Tohoku University, Ibaraki 311-1313, Japan}
\author{G. Knebel}
 \affiliation{Univ. Grenoble Alpes, CEA, Grenoble INP, IRIG, PHELIQS, 38000, Grenoble, France}
\author{W. Knafo}
 \affiliation{Laboratoire National des Champs Magn\'{e}tiques Intenses - EMFL, CNRS, Univ. Grenoble Alpes, INSA-T, Univ. Toulouse 3, 31400 Toulouse, France}
\date{\today}

\begin{abstract}

A structural orthorhombic-to-tetragonal phase transition was recently discovered in the heavy-fermion compound UTe$_2$ at a pressure $p^*\simeq3-8$~GPa [Honda \textit{et al.}, J. Phys. Soc. Jpn. \textbf{92}, 044702 (2023); Huston \textit{et al.}, Phys. Rev. Mat. \textbf{6}, 114801 (2022)]. In the high-pressure tetragonal phase, a phase transition at $T_x=235$~K and a superconducting transition at $T_{sc}=2$~K have been revealed. In this work, we present an electrical-resistivity study of UTe$_2$ in pulsed magnetic fields up to $\mu_0H=58$~T combined with pressures up to $p$ = 6 GPa. The field was applied in a direction tilted by 30~$^\circ$~from \textbf{b} to \textbf{c} in the orthogonal structure, which is identified as the direction \textbf{c} of the tetragonal structure. In the tetragonal phase, the presence of superconductivity is confirmed and signatures of metamagnetic transitions are observed at the fields $\mu_0H_{x1}=24$~T and $\mu_0H_{x2}=34$~T and temperatures smaller than $T_x$. We discuss the effects of uniaxial pressure and we propose that a magnetic ordering drives the transition at $T_x$.
\end{abstract}

\maketitle

\section{Introduction}
\label{Intro}

UTe$_2$ was extensively studied in the past few years [\onlinecite{Aoki2022},\onlinecite{Lewin2023}]. It is orthorhombic and has anisotropic electrical and magnetic properties at ambient pressure [\onlinecite{Ikeda2006,Miyake2019,Eo2022,Ran2019a}]. This heavy-fermion paramagnet [\onlinecite{Paulsen2021}] is superconducting below $T_{sc}=1.5-2.1$~K at ambient pressure (phase SC1) and was presented as a candidate for triplet superconductivity [\onlinecite{Ran2019a},\onlinecite{Aoki2019b}]. A reenforcement of superconductivity under a magnetic field $\mu_0H > 15$~T applied along the direction \textbf{b} [\onlinecite{Knebel2019},\onlinecite{Ran2019b}] arises from the stabilization of a high-magnetic-field superconducting phase (SC2) [\onlinecite{Rosuel2023}]. The presence of a second reentrant superconducting phase (SC-PPM), which is fully decoupled from SC1, was found for a magnetic field tilted by 30~$^\circ$~from \textbf{b} toward \textbf{c} [\onlinecite{Ran2019b,Knafo2021,Schonemann2024,Helm2023,Miyake2021}]. The two superconducting phases SC2 and SC-PPM are closely related to a metamagnetic transition at the field $H_m$ [\onlinecite{Knafo2019},\onlinecite{Miyake2019}]. By considering the three dimensions of field directions, the domain of stability of SC-PPM was further identified as a distorted halo [\onlinecite{Lewin2024},\onlinecite{Wu2024}]. At ambient pressure and zero field, quasi two-dimensional antiferromagnetic (AF) fluctuations have been detected [\onlinecite{Duan2020,Knafo2021b,Butch2022,Tokunaga2019,Tokunaga2022}]. They are gapped in the superconducting phase SC1 [\onlinecite{Duan2021},\onlinecite{Raymond2021}] and are probably involved in the superconducting pairing mechanism [\onlinecite{Monthoux2007}]. The Fermi-liquid electrical-resistivity coefficient $A$ and nuclear-magnetic-resonance (NMR) relaxation rates are peaked at $H_m$, indicating the presence of critical magnetic fluctuations which may play a role for the field-induced superconducting phases [\onlinecite{Knafo2021},\onlinecite{Knafo2019},\onlinecite{Tokunaga2023},\onlinecite{Thebault2022}]. Theoretical works considered the relation between superconductivity and the magnetic fluctuations [\onlinecite{Ishizuka2021,Kreisel2022,Shishidou2021,Tei2024}] and the symmetries of the different superconducting phases [\onlinecite{Nakamine2019},\onlinecite{Nakamine2021}].

Under pressure $p$, SC1 is suppressed with a linear decrease with $p$ of the associated critical temperature, and a second superconducting phase SC2 appears at pressures larger than 0.3~GPa, with a maximal critical temperature $T_{sc} \simeq 3$~K at $p \simeq 1.2$~GPa [\onlinecite{Braithwaite2019,Aoki2020,Thomas2020,Kinjo2023,Wu2024b}]. Similar saturations of the NMR Knight shifts below $T_{sc}$ [\onlinecite{Kinjo2023}] and a boundary between SC1 and SC2 in the low temperature ($p$,$H$) phase diagram by tunnel-diode-oscillator technique [\onlinecite{Lin2020}] have been determined. They suggest that the superconducting phase stabilized under pressure may be the same phase SC2 as that induced by a magnetic field, which needs now to be confirmed by thermodynamic measurements, as heat capacity or thermal expansion, under pressure combined with magnetic field. Magnetic-susceptibility measurements showed a change of the magnetic anisotropy at the critical pressure $p_c \simeq 1.7 $~GPa [\onlinecite{Li2021,Kinjo2022,Kinjo2023,Ambika2022}], beyond which superconductivity disappears and an incommensurate antiferromagnetic order is established [\onlinecite{Braithwaite2019},\onlinecite{Knafo2023}]. Pressure was also combined with magnetic fields in different directions, unveiling additional field-induced superconducting phases [\onlinecite{Knebel2020,Aoki2020,Lin2020,Valiska2021,Valiska2021SI,Ran2021,Aoki2021,Kim2023,Knebel2023}].

\begin{figure}[t]
\includegraphics[width=0.9\columnwidth,trim={0cm 6cm 0cm 3cm},clip]{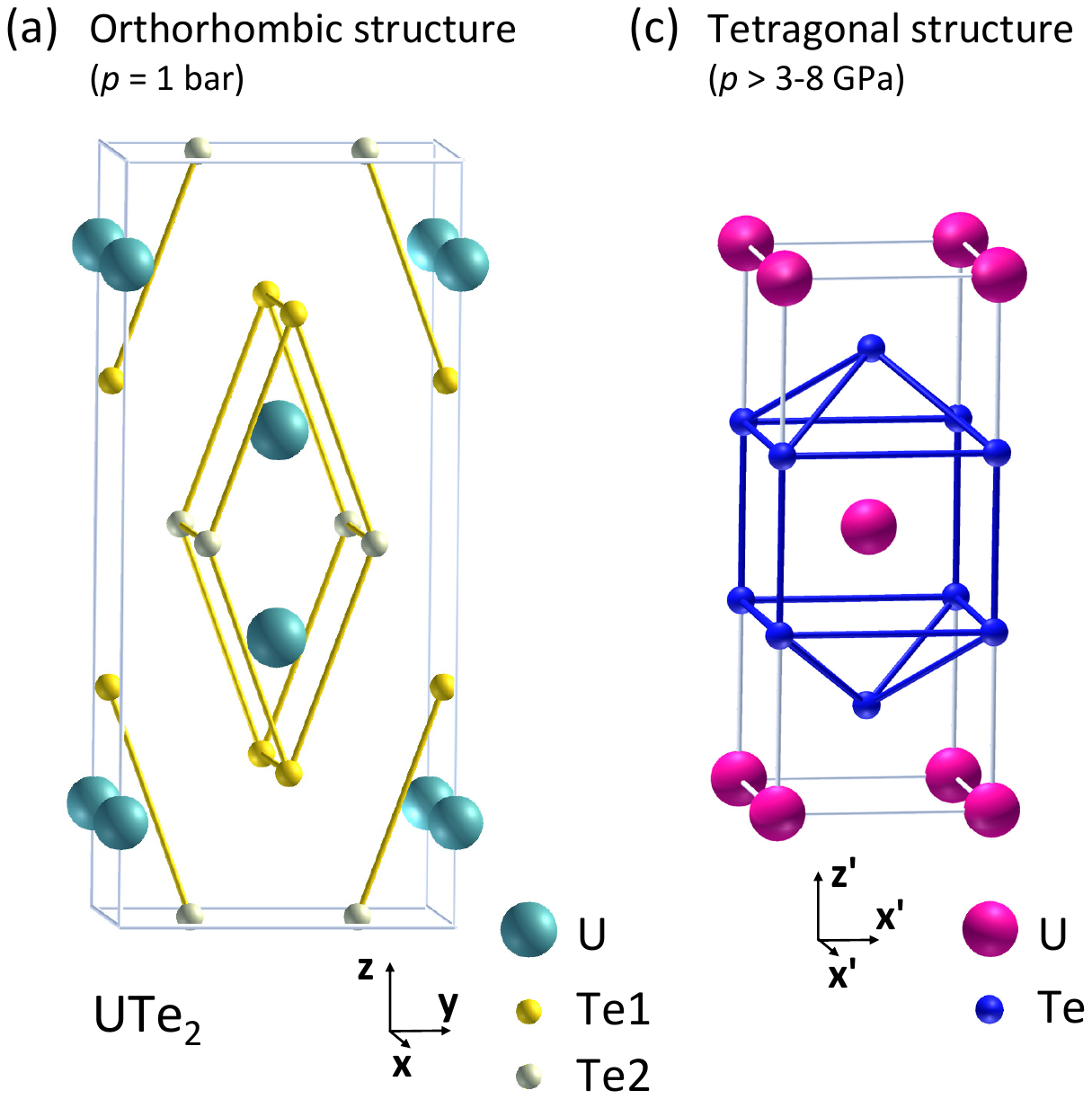}
\caption{\label{Fig1} Unit cells of UTe$_2$ (a) in its low-pressure orthorhombic structure and (b) in its high-pressure tetragonal structure.}
\end{figure}

Recently, a study by electrical-resistivity measurements combined with single-crystal x-ray diffraction under pressures up to $p=10$~GPa showed the presence of a structural transition at $p^*\simeq3-8$~GPa, from a low-pressure orthorhombic structure with the space group $Immm$ into a high-pressure tetragonal structure with the space group $I4/mmm$ [\onlinecite{Honda2023}] (see Figure \ref{Fig1}). The structural transition was confirmed by powder x-ray diffraction studies under pressure up to $p=24$~GPa [\onlinecite{Huston2022}] and 30~GPa [\onlinecite{Deng2024}] and related with a valence change [\onlinecite{Deng2024},\onlinecite{Wilhelm2023}]. A structural transition was also predicted in [\onlinecite{Hu2022}]. In the tetragonal phase, a kink in the electrical resistivity indicates a phase transition at the temperature $T_x \simeq 235 $~K [\onlinecite{Honda2023,Valiska2021,Valiska2021SI}] and a superconducting phase develops below $T_{sc}=2$~K [\onlinecite{Honda2023}]. A transition under a magnetic field was also seen in UTe$_2$ in its high-pressure tetragonal phase at temperatures below $T_x$, and its trace at around 30~T was lost at low temperature [\onlinecite{Valiska2021},\onlinecite{Valiska2021SI}]. However, in that study the uncertainty in the magnetic-field direction \textbf{H} $\simeq||$\textbf{b} was large, with an estimated misorientation of $\simeq15-30$~$^\circ$, due to a sample positioning issue in the pressure cell (see discussion in the Supplemental Material [\onlinecite{SM}]).

Here we present electrical-resistivity measurements performed on a UTe$_2$ single crystal under pressure up to $p=  6$~GPa and magnetic field up to $\mu_0H \simeq 58 $~T. The magnetic field was applied along a direction tilted by $30$~$^\circ$~from \textbf{b} toward \textbf{c} in the orthorhombic structure. Signatures of the tetragonal phase are found for $p\ge3.1$~GPa. Magnetic-field-temperature phase diagrams are constructed for the different pressures. The Fermi-liquid coefficient $A$ is extracted from measurements made in the orthorhombic phase. In the tetragonal phase, we confirm the presence of superconductivity and metamagnetic transitions with a large hysteresis are observed at $\mu_0H_{x1}=24 $~T and $\mu_0H_{x2}=34$~T.

\section{Methods}
\label{Methods}

The UTe$_2$ single crystal measured here was grown by the molten-salt-flux method [\onlinecite{Sakai2022}]. Its electrical resistivity $\rho_{xx}$ was measured with an electrical current along $\mathbf{a}$ under magnetic fields $\mu_0\mathbf{H}$ up to 58~T tilted by 30~$^\circ$ from $\mathbf{b}$ toward $\mathbf{c}$. The field was combined with pressures up to 6~GPa and temperatures down to 1.4~K. The sample was polished with a tilt by an angle of 7~$^\circ$ from a cleaving surface normal to the direction $\mathbf{n}$ of Miller indices $(0,1,1)$ (see [\onlinecite{Knafo2021b}]), ending in two faces normal to a direction tilted by 30~$^\circ$ from $\mathbf{b}$ towards $\mathbf{c}$. The orientation of the crystal was ensured by Laue diffraction after the polishing process. Its electrical resistivity was measured using the four-contact technique, with an excitation current of 4~mA at a frequency of $\simeq 50 $~kHz. Pulsed-magnetic-field experiments were performed using long-duration 60-T magnets at the LNCMI-Toulouse. Here, we used a Bridgman-type cell specifically designed for the pulsed magnetic fields [\onlinecite{Braithwaite2016}]. To avoid substantial heating by eddy currents, ceramic anvils and a pyrophyllite gasket were used in the cell, whose body is made of MP35N. Most of the data presented here correspond to the rise of the field pulses, where the heating of the sample due to eddy currents in the cell is estimated, for 58-T pulses, to be less than 0.1~K at temperatures $T \le4.2 $~K, close to $1-1.5$~K at temperatures from 5~K to 20~K and negligible at higher temperatures. For the falling-field data considered in this manuscript, the heating effects were evaluated during the pulse and are detailed in the Supplemental Material [\onlinecite{SM}]. For low-field pulses ($\mu_0H \le 30$~T), heating effects were negligible. Pressure was changed at room temperature and was estimated by measuring the superconducting transition temperature of a lead gauge. More information about the pressure cell set-up can be found in two technical papers [\onlinecite{Braithwaite2016},\onlinecite{Settai2015}].

\section{Results}
\label{Results}

\subsection{Overview}

\begin{figure}[t]
\includegraphics[width=0.95\columnwidth]{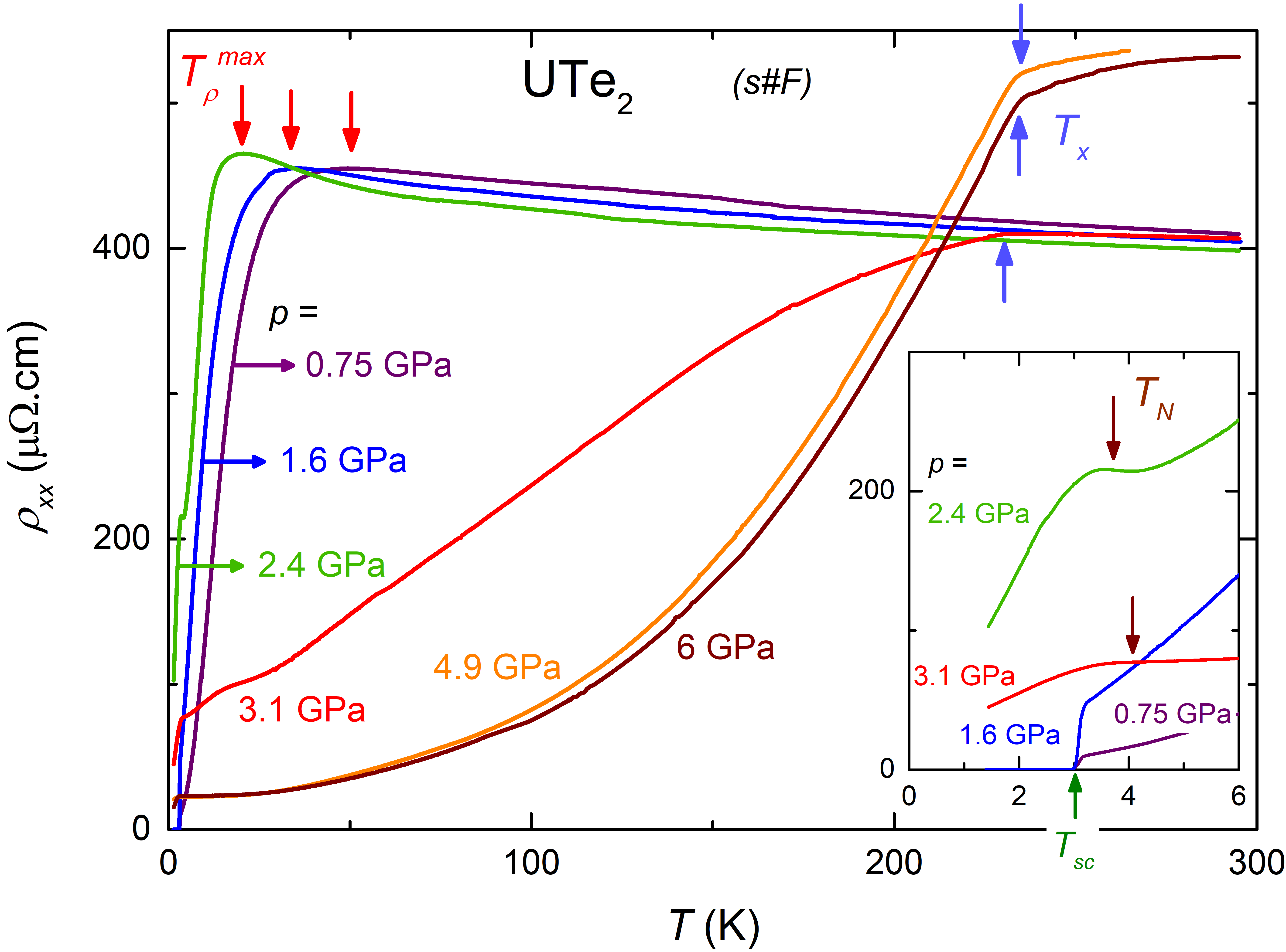}
\caption{\label{Fig2} Temperature dependence of the electrical resistivity $\rho_{xx}$, measured with $\mathbf{I}\parallel\mathbf{a}$, of UTe$_2$ under different hydrostatic pressures and zero magnetic field. Inset: zoom on the onset of superconductivity and antiferromagnetic order at low temperature.}
\end{figure}

Figure \ref{Fig2} compares the temperature-dependence of the electrical resistivity $\rho_{xx}$ for different pressures $p$ from 0.75 to 6~GPa. At $p = 0.75$ and 1.6~GPa, superconductivity develops below $T_{sc}\simeq3$~K. The characteristic temperature $T_\rho^{max}$ of the maximum in resistivity is decreasing with increasing pressure from $p = 0.75 $~GPa to 2.4~GPa. We see the onset of the antiferromagnetic order at the N\'{e}el temperature $T_N = 3.7 $~K for the pressures $p = 2.4 $~GPa and $p = 3.1$~GPa, i.e., above the critical pressure $p_c \simeq 1.7 $~GPa. This AF order is identified as a signature of the orthorhombic phase [\onlinecite{Knafo2023}]. A kink at $T_x \simeq235$~K is visible in the electrical resistivity measured for $p\ge3.1$~GPa. Such kink was also observed in the tetragonal phase in [\onlinecite{Honda2023}] and, in the following, we will identify it as a signature of the tetragonal phase. At $p = 3.1 $~GPa, signatures of both $T_N$ and $T_x$ indicate the coexistence of domains with the orthorhombic and tetragonal phases. At $p$ = 4.9 GPa and 6 GPa, the onset of superconductivity in the tetragonal phase is identified at $\simeq 3 $~K and the electrical resistivity increases monotonously with temperature up to $T_x$. At $p = 6 $~GPa, superconductivity with zero resistivity is observed at temperatures below $T_{sc} = 1.45 $~K.

\begin{figure}[t]
\includegraphics[width=\columnwidth]{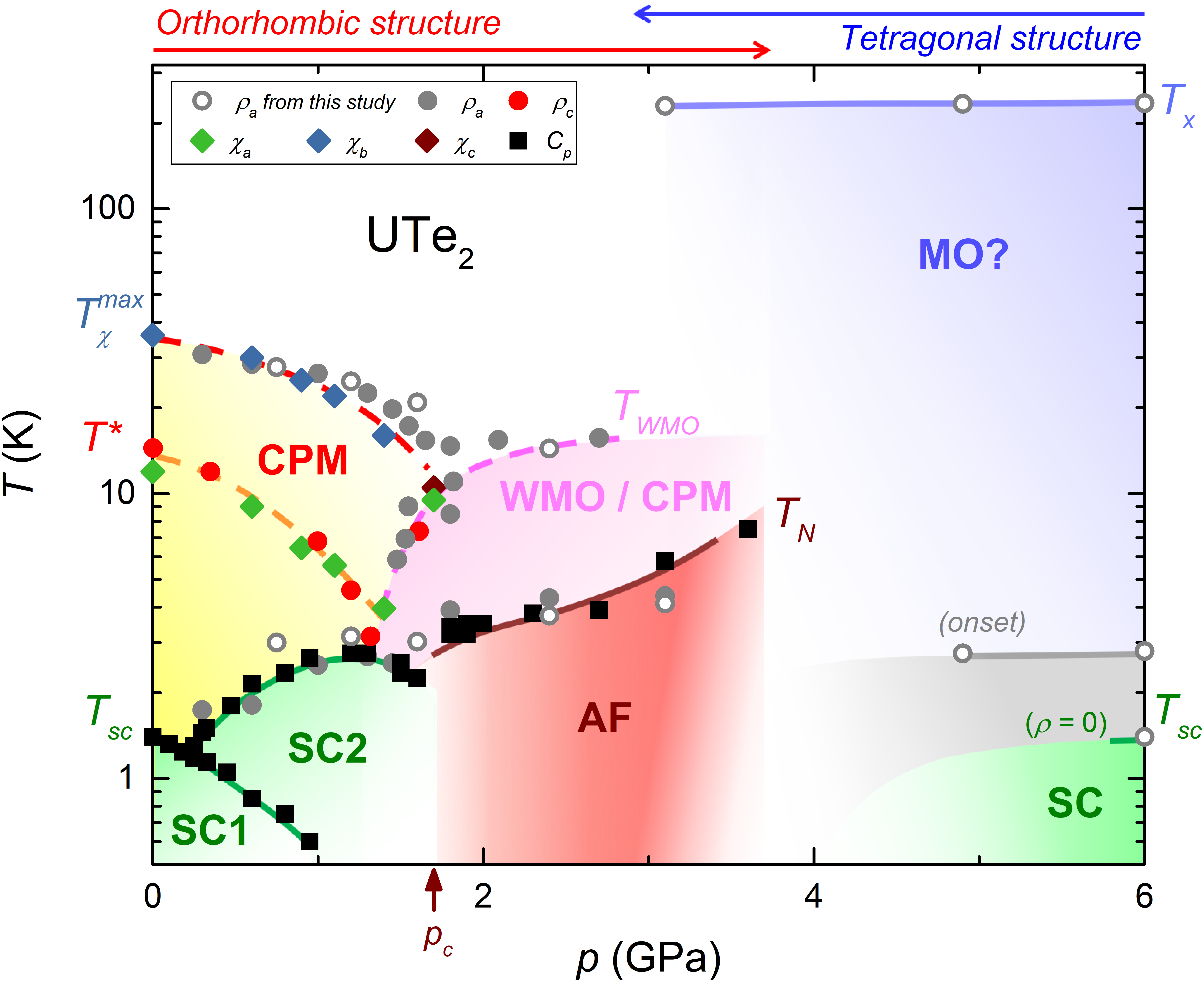}
\caption{\label{Fig3} Pressure-temperature phase diagram of UTe$_2$ obtained from data published in [\onlinecite{Braithwaite2019,Li2021,Valiska2021,Knebel2023}] and data collected here. CPM denotes the correlated paramagnetic regime, WMO/CPM the weak magnetic order / correlated paramagnetic regime, SC1, SC2 and SC the different superconducting phases, AF the antiferromagnetic phase and MO? a suspected magnetically-ordered phase.}
\end{figure}

Figure \ref{Fig3} presents the pressure-temperature phase diagram obtained from data collected here and data published in [\onlinecite{Braithwaite2019},\onlinecite{Li2021},\onlinecite{Valiska2021},\onlinecite{Knebel2023}]. The low-pressure region exhibits two crossovers at $T^{max}_\chi \simeq 35$~K and $T^* \simeq 15 $~K and two different superconducting phases SC1 and SC2. The first crossover at $T^{max}_\chi$ is associated with a maximum in the magnetic susceptibility measured in a magnetic field along \textbf{b} [\onlinecite{Ikeda2006,Miyake2019,Li2021}] and with a maximum in $\Delta\rho_{xx}$ obtained after a background subtraction to the electrical resistivity $\rho_{xx}$ [\onlinecite{Valiska2021}]. It delimits a correlated paramagnetic (CPM) regime. The second  crossover at $T^* \simeq 15$~K is associated with a kink in the magnetic susceptibility measured in a magnetic field along \textbf{a} and a maximum in the electrical resistivity $\Delta\rho_{zz}$ (here subtracting a background has little effect due to the smaller temperature scale) [\onlinecite{Thebault2022},\onlinecite{Knebel2023}]. The two crossover temperatures decrease with increasing pressure up to the critical pressure $p_c \simeq$ 1.7 GPa. Above this critical pressure, antiferromagnetic order is established below $T_N$ $\simeq3.5$~K (defined here at the inflection point of $\rho_{xx}$ versus $T$) and a crossover is observed at $T_{WMO} \simeq$ 8-12 K in electrical-resistivity (defined at a kink near $p_c$ or a maximum much beyond $p_c$ of $\rho_{xx}$ versus $T$) [\onlinecite{Valiska2021},\onlinecite{Valiska2021SI},\onlinecite{SM}] and magnetic-susceptibility data [\onlinecite{Li2021}]. $T_{WMO}$ is the onset of a weak-magnetic-order regime, which is delimited by a maximum in the magnetic susceptibility and by a metamagnetic transition, and can be also identified as a CPM regime (see [\onlinecite{Valiska2021}]). This regime is noted WMO/CPM here. At $p = 3.1 $~GPa, signatures of domains with the orthorhombic and tetragonal structures are observed in our data. By further increasing the pressure, the tetragonal phase is fully stabilized. The orthorhombic antiferromagnetic phase is no longer visible and the onset of the tetragonal superconducting phase is observed. Differences between the critical pressure $p^*\lesssim3$~GPa of the structural transition determined using the same pressure cells here and in [\onlinecite{Valiska2021}] and the large set of critical pressures $p^*\simeq3-8$~GPa reported in [\onlinecite{Honda2023},\onlinecite{Huston2022}] may be due to non-hydrostatic conditions of pressure (see discussion in Section \ref{SectionIVC}).

\begin{figure}[t]
\includegraphics[width=\columnwidth]{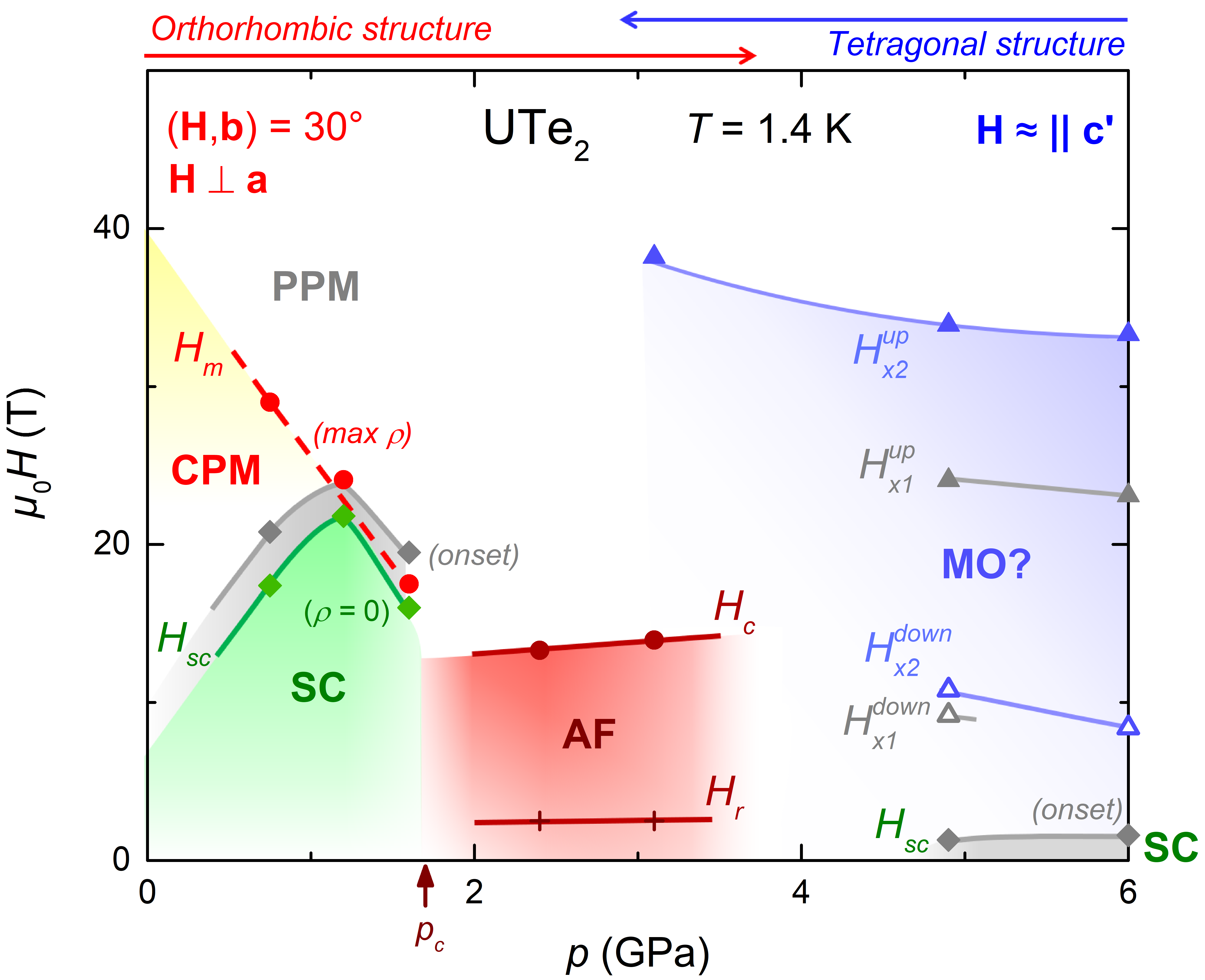}
\caption{\label{Fig4} Pressure-magnetic-field phase diagram of UTe$_2$ at $T = 1.4 $~K and under a magnetic field tilted by 30~$^\circ$ from $\mathbf{b}$ toward $\mathbf{c}$ (low-pressure orthorhombic structure) or along $\simeq\mathbf{c}'$ (high-pressure tetragonal structure). CPM denotes the correlated paramagnetic regime, PPM the polarized paramagnetic regime, SC superconductivity, AF the antiferromagnetic phase and MO? a suspected magnetically-ordered phase.}
\end{figure}

\begin{figure*}[t]
\includegraphics[width=0.85\textwidth]{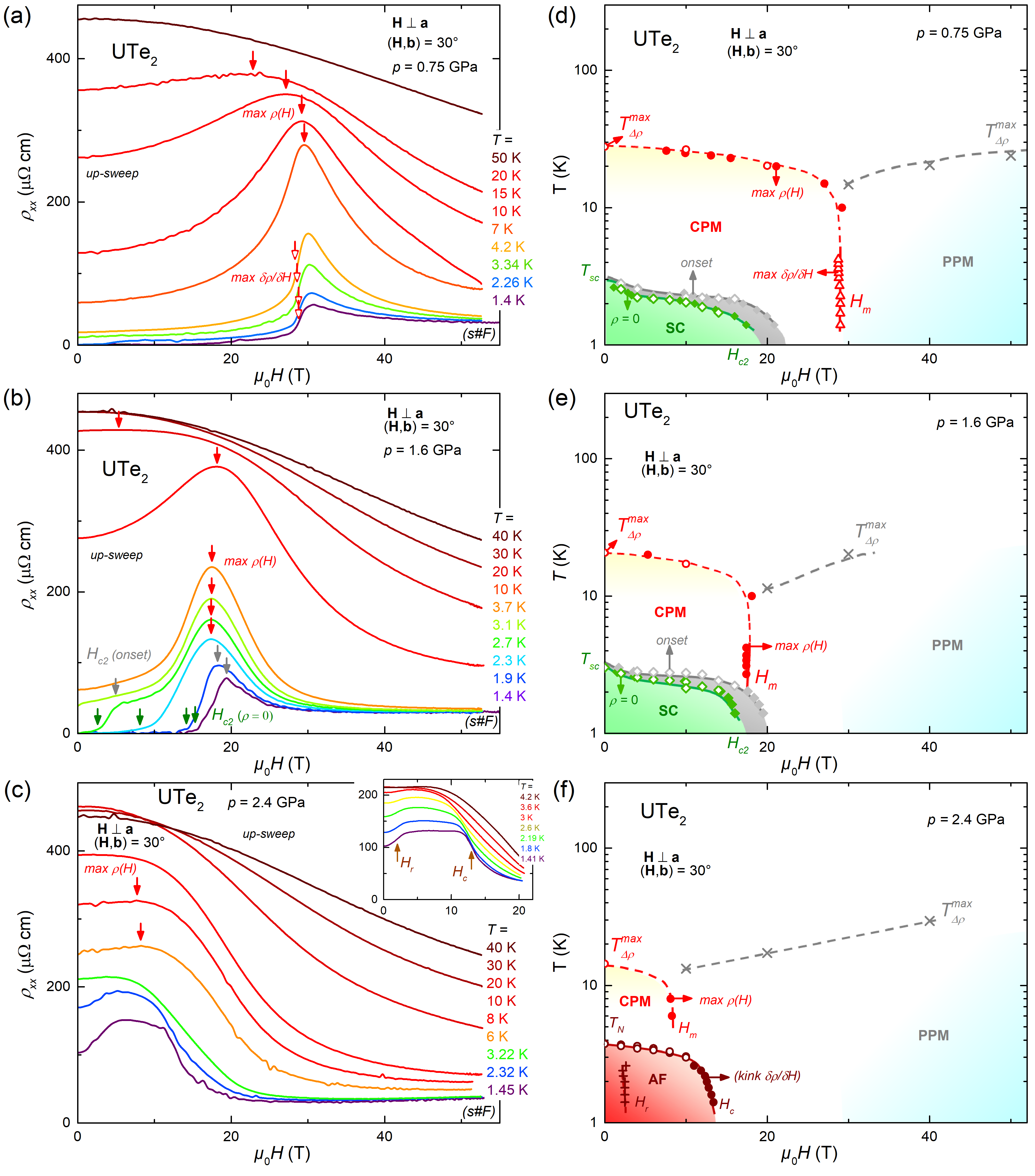}
\caption{\label{Fig5} Left-hand graphs: Electrical resistivity versus magnetic field at temperatures between 1.4~K and 50~K and pressures (a) $p=0.75$~GPa, (b) $p=1.6$~GPa, and (c) $p=2.4$~GPa. The data were obtained during the rise of the magnetic field. Right-hand graphs: Magnetic-field-temperature phase diagrams at (d) $p=0.75$~GPa, (e) $p=1.6$~GPa, and (f) $p=2.4$~GPa. CPM denotes the correlated paramagnetic regime, PPM denotes the polarized paramagnetic regime, SC superconductivity, and AF the antiferromagnetic phase. In (d-f), open symbols correspond to points extracted from $\rho(T)$ data and closed symbols correspond to points extracted from $\rho(H)$ data.}
\end{figure*}

Figure \ref{Fig4} shows the pressure-magnetic-field phase diagram obtained here at $T \simeq 1.4 $~K. Details about the data will be given in Sections \ref{SectionIIIA} and \ref{SectionIIIB}. The magnetic field is tilted by 30~$^\circ$ from \textbf{b} toward \textbf{c} in the orthorhombic phase. We will see in Section \ref{SectionIVB} that this field direction is identified as the direction \textbf{c}' of the tetragonal phase. In the orthorhombic phase, superconductivity is enhanced under pressure and is delimited by the metamagnetic transition for 1.1~GPa~$\le p \le p_c$. The metamagnetic field $H_m$ weakens when the pressure is increased and no signature of $H_m$ is found above the critical pressure $p_c \simeq 1.7 $~GPa. For $p > p_c$, an antiferromagnetic order is stabilized [\onlinecite{Knafo2023}] and its phase is delimited by the critical field $H_c$. For $p\ge4.9$~GPa, no signature of the electronic properties in the orthorhombic structure is found and two transitions at the fields $H_{x1}$ and $H_{x2}$ are observed with a large hysteresis. $H_{xi}^{up}$ and $H_{xi}^{down}$ (for $i=1,2$) denote respectively the transition field observed during the rise and fall of the magnetic field. At $p = 6 $~GPa, the boundary of the superconducting phase is observed at $\mu_0H_{c2}\simeq1.5 $~T. At $p=3.1$~GPa, the transitions at $\mu_0H_c=13.5$~T and $\mu_0H_{x2}=38$~T are identified, implying a probable coexistence of orthorhombic and tetragonal domains.

\subsection{Orthorhombic phase}
\label{SectionIIIA}

\begin{figure}[t]
\includegraphics[width=0.85\columnwidth]{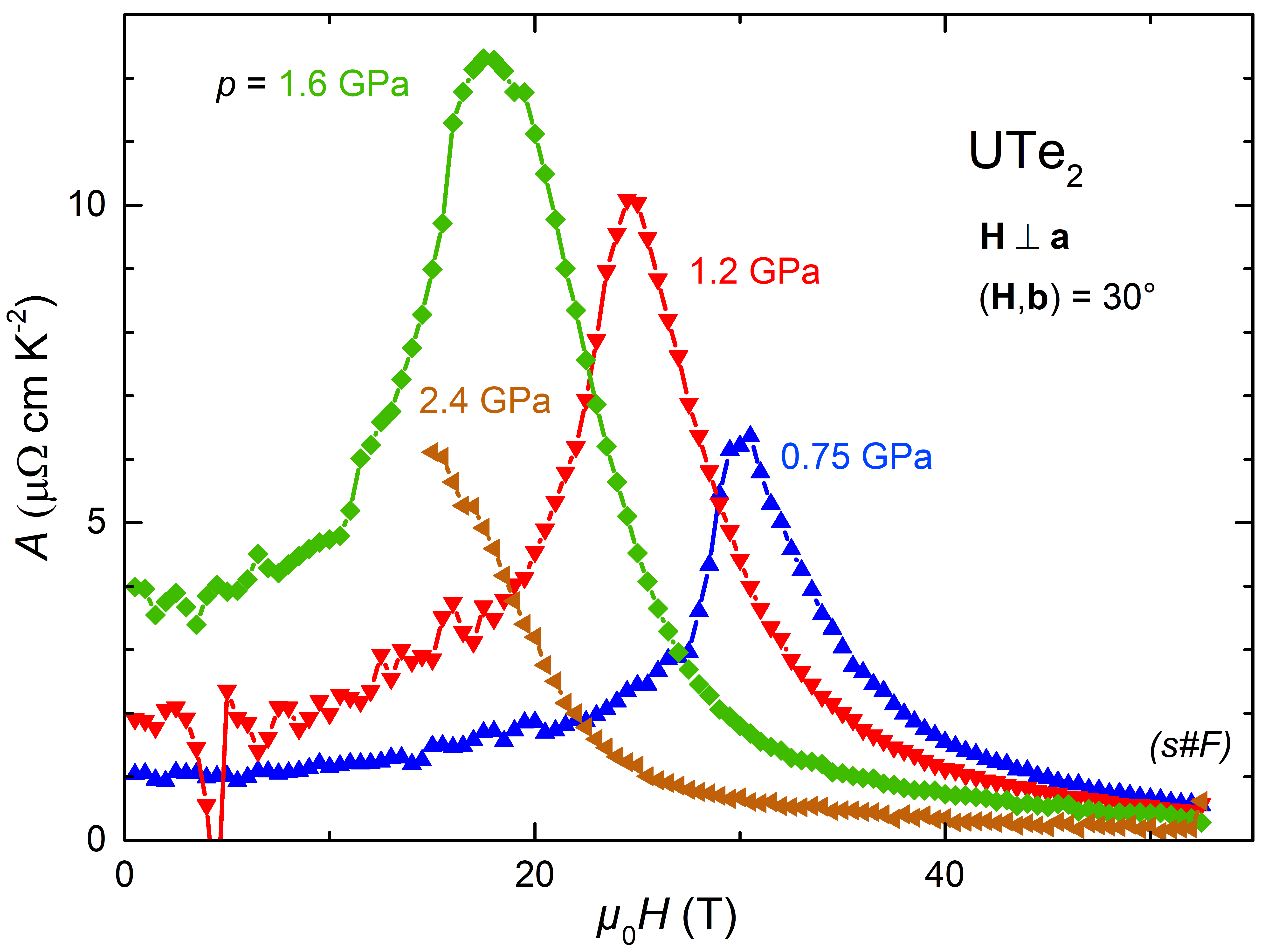}
\caption{\label{Fig6} Field-dependence of the Fermi-liquid electrical-resistivity coefficient $A$ of UTe$_2$ at different pressures up to 2.4~GPa  in a magnetic field tilted by 30~$^\circ$ from $\mathbf{b}$ toward $\mathbf{c}$.}
\end{figure}

Figures \ref{Fig5}(a-c) present electrical-resistivity data obtained at pressures from $p = 0.75 $~GPa to 2.4~GPa for temperatures ranging from $T = 1.4$~K to 50~K. Details about the definition of the transition and crossover fields are given in the Supplemental Material [\onlinecite{SM}]. At $p = 0.75 $~GPa and at low temperatures, a step-like transition in the resistivity is the signature of the metamagnetic transition at $\mu_0H_m = 39 $~T. At higher temperatures, this transition turns into a broad maximum. At $p = 1.6 $~GPa, the signature of the metamagnetic transition is masked by the onset of superconductivity at low temperature. At $p = 2.4 $~GPa, the antiferromagnetic order is stabilized below $T_N = 3.7 $~K. At temperatures $T<T_N$, kinks in the electrical resistivity are observed at the magnetic fields $\mu_0H_r$ and $\mu_0H_c$, which reach 2.5~T and 13.5~T, respectively, at low temperature (see also Ref. [\onlinecite{Valiska2021}]). Figures \ref{Fig5}(d-f) show the magnetic-field-temperature phase diagrams obtained from our resistivity data at pressures $p = 0.75 $~GPa, 1.6~GPa and 2.4~GPa, respectively. At $p = 0.75 $~GPa and 1.6~GPa, $H_{c2}$ exhibits a change of curvature at $T\simeq2.4$~K, whose origin is not understood so far. The low-temperature superconducting field $\mu_0H_{c2}$ reaches $\simeq17$~T at $p=1.6$~GPa, i.e., near the critical pressure, here for $\mathbf{H}$ tilted by 30~$^\circ$~from $\mathbf{b}$ towards $\mathbf{c}$, which is strongly enhanced in comparison with that of $\simeq6$~T found for $\mathbf{H}\parallel\mathbf{b}$ [\onlinecite{Knebel2020}]. The fields and temperatures, respectively defined at the broad maxima in $\rho_{xx}(H)$ and $\Delta\rho_{xx}(T)$, which was obtained after a background subtraction as done in previous studies [\onlinecite{Thebault2022},\onlinecite{Valiska2021}], delimit the CPM regime (see details in the Supplemental Material [\onlinecite{SM}]). At $p = 2.4 $~GPa, the antiferromagnetic phase is delimited by $H_c$ and no trace of superconductivity is found at low temperature.

Figure \ref{Fig6} shows the field dependence of the Fermi-liquid coefficient $A$ determined under pressures up to $p =2.4$~GPa. The fits by $\rho=\rho_0+AT^2$ to the data used to extract $A$ are provided in the Supplemental Material [\onlinecite{SM}]. At the pressures $p = 0.75 $~GPa, 1.2~GPa, and 1.6~GPa, the coefficient $A$ exhibits a maximum at $H_m$. At $p=2.4$~GPa, the electrical resistivity does not follow a quadratic temperature dependence in the antiferromagnetic phase for $H<H_c$. A decrease of $A$ with an increasing magnetic field is observed in the polarized paramagnetic regime reached for $H>H_c$. For $p\ge3.1$~GPa, the electrical resistivity does not follow a $T^2$ behavior at all fields investigated here.

\begin{figure}[t]
\includegraphics[width=0.9\columnwidth]{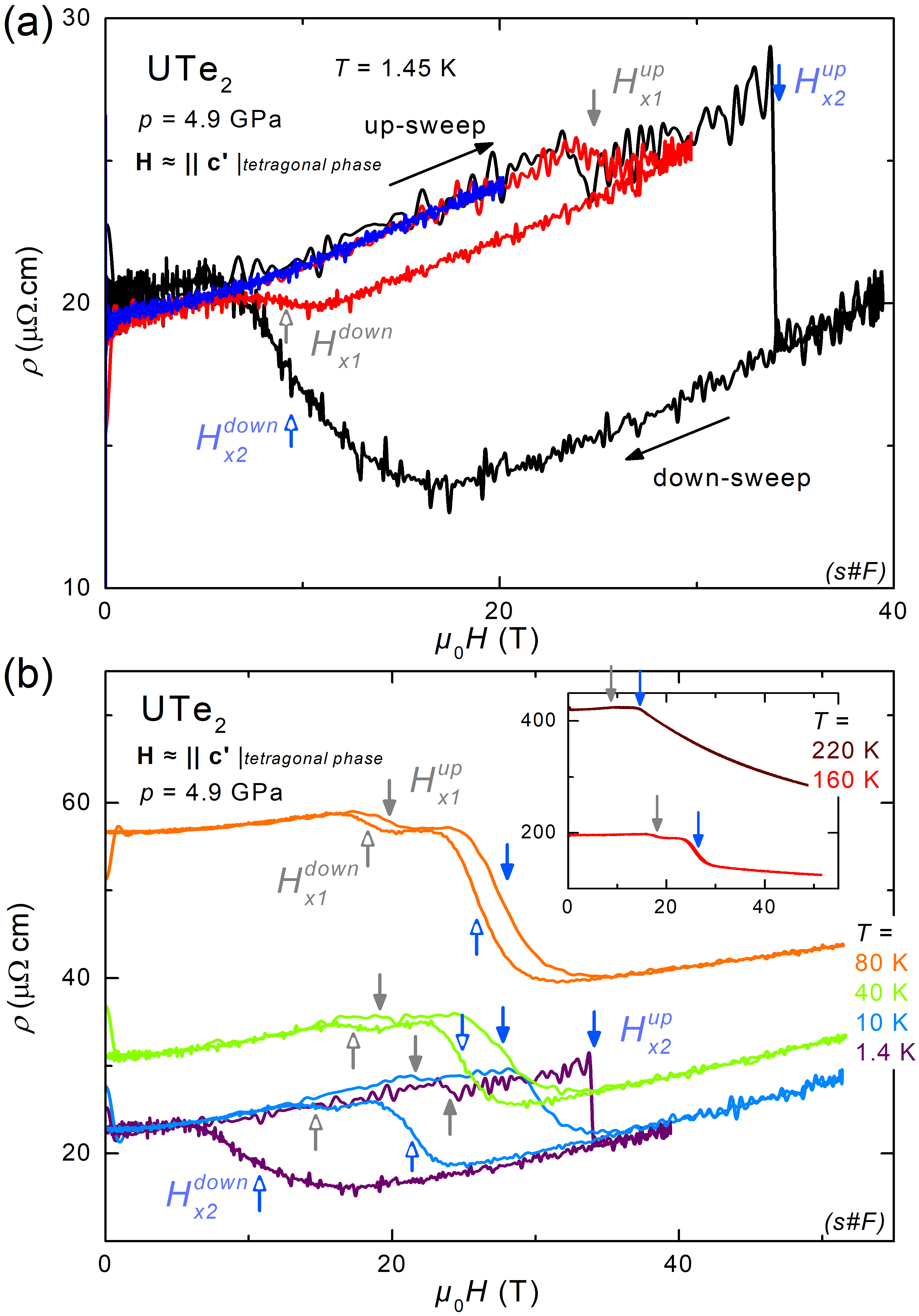}
\caption{\label{Fig7} (a) Magnetic-field dependence of the electrical resistivity $\rho_{xx}$ of UTe$_2$ at $T=1.4$~K and $p=4.9$~GPa obtained from pulsed-field shots of different strengths. (b) Magnetic-field dependence of the electrical resistivity $\rho_{xx}$ of UTe$_2$ at $p=$4.9~GPa and temperatures varying from 1.4~K to 220~K. Data obtained during the rise and fall of the pulse are shown here. Closed arrows indicate the transition fields in the rise of the field and open arrows indicate the transition fields in the fall of the field.}
\end{figure}

\begin{figure*}[ht]
\includegraphics[width=0.85\textwidth]{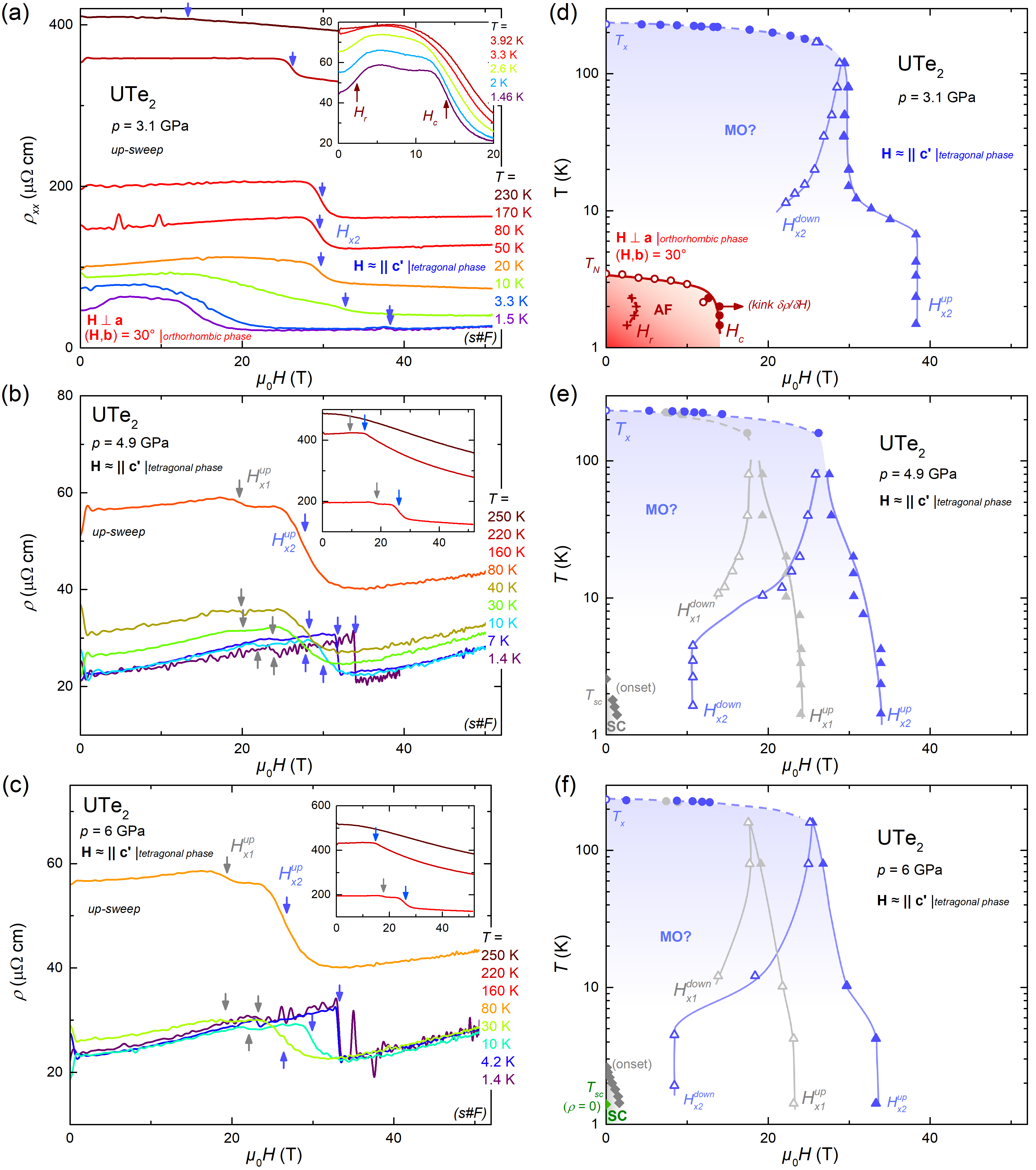}
\caption{\label{Fig8} Left-hand graphs: Electrical resistivity versus magnetic field at temperatures between 1.4~K and 250~K and pressures (a) $p=3.1$~GPa, (b) $p=4.9$~GPa, and (c) $p=6$~GPa. The data were obtained during the rise of the magnetic field. Right-hand graphs: Magnetic-field-temperature phase diagrams at (d) $p=3.1$~GPa, (e) $p=4.9$~GPa, and (f) $p=6$~GPa. SC denote superconductivity, AF the antiferromagnetic phase, and MO? a suspected magnetically-ordered phase.}
\end{figure*}

\begin{figure}[t]
\includegraphics[width=0.85\columnwidth]{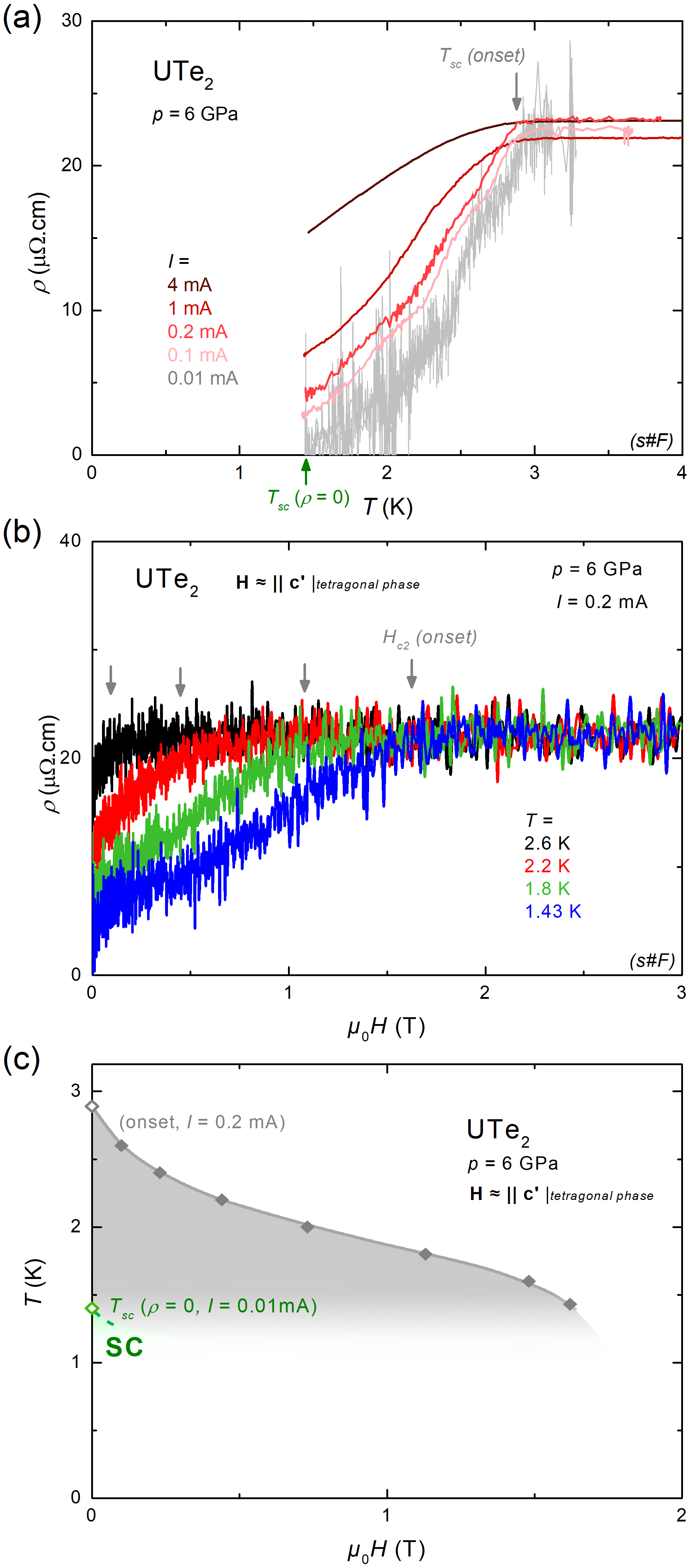}
\caption{\label{Fig9} (a) Temperature dependence of the electrical resistivity $\rho_{xx}$ of UTe$_2$ at $p=6$~Gpa and zero magnetic field, measured with different electrical currents $I$ varying from 0.01~mA to 4~mA. (b) Magnetic-field dependence of the electrical resistivity $\rho_{xx}$ measured with a current $I=0.2$~mA, at temperatures from 1.43~K to 2.6~K and at $p=6$~Gpa. (c) Magnetic-field-temperature phase diagram of the superconducting phase at $p=6$~Gpa.}
\end{figure}

\subsection{Tetragonal phase}
\label{SectionIIIB}

Figure \ref{Fig7}(a) shows the electrical resistivity measured at $p = 4.9 $~GPa and $T = 1.45 $~K for the rise and fall of pulsed magnetic fields of different strengths, from 20~T to 40~T. At $p = 4.9$~GPa, UTe$_2$ is in its tetragonal phase and the magnetic field is applied near to the direction \textbf{c}' of the tetragonal structure (see Section \ref{SectionIVB}). A transition with a step-like variation of $\rho$ by $\simeq2$~$\mu\Omega$cm occurs at $\mu_0H_{x1}^{up}=24$~T during the rise of the magnetic field. This transition has a strong hysteresis and is characterized by the field $\mu_0H_{x1}^{down}=9$~T during the fall of the magnetic field. A second transition with a larger step-like variation of $\rho$, by $\simeq10$~$\mu\Omega$cm, occurs at $\mu_0H_{x2}^{up}=34$~T in the rising fields. This second transition has also a large hysteresis and is characterized by the field $\mu_0H_{x2}^{down}=9.5$~T in the falling fields. $H_{x2}^{down}$ and $H_{x1}^{down}$ cannot be distinguished for the most intense magnetic field pulses (up to more than 34~T). The initial state is recovered at the end of the magnetic field pulse for $H < H_{x1}^{down}$. We note that, due to eddy currents generated in the pressure cell, heating effects during the 42-T magnetic field pulses lead to a temperature increase on the sample estimated by $\Delta T=0.1 $~K at the maximum field and $\Delta T=0.6$~K at the end of the pulse, which has a small incidence on the data presented here. Figure \ref{Fig7}(b) presents the electrical resistivity for temperatures varying from 1.4~K to 220~K measured under a magnetic field up to 60~T, at the pressure $p=4.9$~GPa. By increasing the temperature, the signatures at the two transitions become broader, their hysteresis weakens and disappears at temperatures higher than 160~K. At temperatures $T\ge10$~K, the characteristic fields $H_{x1}^{down}$ and $H_{x2}^{down}$ can be easily distinguished from each other.

Figures \ref{Fig8}(a-c) present the field-dependence of the electrical resistivity measured at pressures $p$ from 3.1~GPa to 6~GPa, from the rises of the field pulses. At $p=3.1$~GPa, the low-temperature electrical resistivity shows kinks at the fields $H_r$ and $H_c$, which are characteristic of the orthorhombic antiferromagnetic phase, and a step driven by a transition at the field $H_{x2}$, which is characteristic of the tetragonal phase. The anomaly at $H_{x2}$ weakens at high temperature and only a kink is visible at $T=230$~K. At $p=4.9$~GPa and 6~GPa, two transitions are observed at the fields $H_{x1}$ and $H_{x2}$, at temperatures from $T=1.4$~K to 235~K, and no signature of the orthorhombic antiferromagnetic phase is found. No signatures of $H_{x1}$ and $H_{x2}$ are observed at temperatures larger than $T_x=235$~K. Figures \ref{Fig8}(d-f) present the magnetic-field-temperature phase diagrams obtained from the electrical-resistivity data. At $p=3.1$~GPa, signatures of both the tetragonal high-temperature phase delimited by $T_x = 235 $~K and $\mu_0H_{x2}^{up}=38$~T and the orthorhombic antiferromagnetic order established at temperatures below $T_N$ are observed. This indicates that the crystal may be inhomogeneous, with the presence of orthorhombic and tetragonal domains. The transition at $H_{x2}$ exhibits a strong hysteresis at $T=10$~K, which weakens with increasing temperature. At $p=4.9$~GPa, the absence of the orthorhombic AF phase allows to see the hysteresis of the two transitions at $H_{x1}$ and $H_{x2}$ at low temperatures. The temperature evolution of these two transitions is similar, their hystereses end at $T\simeq120$~K and their signatures disappear at temperatures above $T_x$. A temperature correction was applied for the $H_{x2}^{down}$ data shown at temperatures $T < 20 $~K in these phase diagrams (see Supplemental Material [\onlinecite{SM}]). The data collected at $p=6$~GPa are very similar to the data collected at $p=4.9$~GPa, but superconductivity with $\rho=0$ is only observed at $p=6$~GPa (in data collected at zero field or from low-field pulses, with a small electrical current, see next paragraph). No noticeable difference between the signatures of the metamagnetic transitions at the two pressures is found.

Figure \ref{Fig9}(a) presents the temperature dependence of the electrical resistivity of UTe$_2$, measured for different electrical currents, in zero field and at the pressure $p=6$~GPa. With an electrical current $I=0.01$~mA, superconductivity with zero resistivity occurs below the temperature $T_{sc}=1.45$~K. For higher excitation currents, zero resistivity is not reached but the onset of superconductivity can be defined at a temperature of $\simeq2.9$~K. A large temperature width, of $\simeq1.5$~K, of the superconducting transition is found. Smooth step-like anomalies in the electrical resistivity at temperatures between $T_{sc}$ and 2.9 K indicate an inhomogeneous sample quality, possibly due to the non-hydrostatic pressure generated by our cell. Figure \ref{Fig9}(b) shows the electrical resistivity measured with a current $I=0.2$~mA under a pressure $p=6$~GPa and magnetic fields up to $\mu_0H=3$~T. This current was chosen to extract the superconducting phase boundary under magnetic field with a good signal-over-noise ratio. However, it is much larger than the current of 0.01~mA at which zero resistivity was extracted below $T_{sc}$. This makes the characterization of $H_{c2}$ defined with $\rho=0$ not possible and only the magnetic field at the onset of superconductivity can be extracted. We note that superconductivity could not be evidenced from the high-field-pulses data presented in Figures \ref{Fig7}(b-c), due to the large current $I=4$~mA used but also to the difficulty to extract the field-variation of $\rho$ at the very-beginning of the pulses. The magnetic-field-temperature phase diagram in Figure \ref{Fig9}(c) shows the boundaries of the superconducting phase stabilized at $p=6$~GPa in a magnetic field applied along the direction $\mathbf{c}'$ of the tetragonal structure (see Section \ref{SectionIVB}). This phase diagram is similar to that reported in [\onlinecite{Honda2023}] for a magnetic field applied along the direction $\mathbf{c}$ of the orthorhombic structure, which can be identified as a direction titled in the ($\mathbf{a}'$,$\mathbf{c}'$) plane of the tetragonal structure (see Section \ref{SectionIVB}).

\section{Discussion}
\label{SectionIV}

The present work, in which UTe$_2$ was investigated under pressures up to 6~GPa combined with magnetic fields tilted by 30~$^\circ$ from $\mathbf{b}$ towards $\mathbf{c}$, completes a former study of UTe$_2$ under pressures up to 4~GPa combined with magnetic fields $\mathbf{H}\simeq\parallel\mathbf{b}$, but with a $15-30^\circ$ misorientation of the sample [\onlinecite{Valiska2021},\onlinecite{Valiska2021SI}]. Interestingly, similar pressure variations of the critical field $H_m$ and of the Fermi-liquid coefficient $A$ have been found in the two studies, and we suspect now that the field direction in [\onlinecite{Valiska2021},\onlinecite{Valiska2021SI}] was probably near to that applied here (see discussion in the Supplemental Material [\onlinecite{SM}]). While the two sets of data are quite similar in the low-pressure orthorhombic phase, new information is collected here concerning the high-pressure tetragonal phase. Stronger field-induced anomalies are observed at the lowest temperature ($T=1.4$~K), presumably due to an increase of the sample quality (the sample studied here was grown by the molten-salt-flux method [\onlinecite{Sakai2022}], while the samples studied in [\onlinecite{Valiska2021},\onlinecite{Valiska2021SI}] were grown by the chemical-vapor-transport method [\onlinecite{Aoki2019b}]). A high-pressure and purely-tetragonal phase with signatures of superconductivity is evidenced, and two field-induced transitions (instead of one in [\onlinecite{Valiska2021},\onlinecite{Valiska2021SI}]) are found. A phase transition at $T_x=235$~K under high pressure was evidenced for the first time in our previous work [\onlinecite{Valiska2021},\onlinecite{Valiska2021SI}]. It was later confirmed in [\onlinecite{Honda2023}], where it was identified as a signature of a high-pressure tetragonal phase beyond $p^*=3-8$~GPa (see also [\onlinecite{Huston2022},\onlinecite{Deng2024}]). Concerning the low-pressure orthorhombic phase, incommensurate antiferromagnetism was recently evidenced under pressure beyond $p_c=1.7$~GPa [\onlinecite{Knafo2023}]. In the light of these recent findings, we discuss below our new experimental data: the high-field properties of UTe$_2$ in its orthorhombic phase are considered in Section \ref{SectionIVA} and the high-field properties of UTe$_2$ in its tetragonal phase are considered in Sections \ref{SectionIVB}, \ref{SectionIVC} and \ref{SectionIVD}.

\begin{figure*}[t]
\includegraphics[width=1\textwidth,trim={1cm 2.5cm 1cm 0cm},clip]{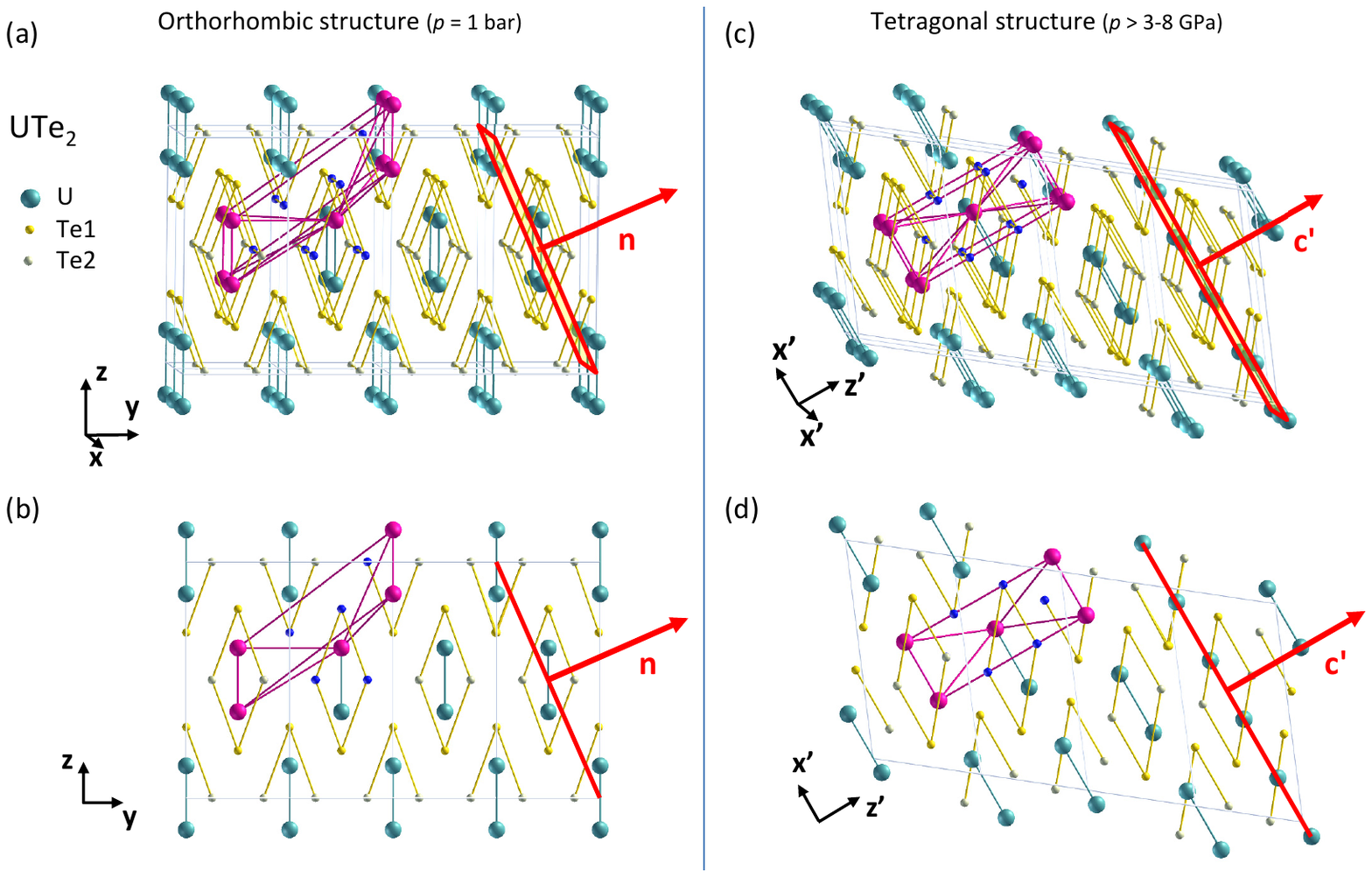}
\caption{\label{Fig10} Views of UTe$_2$ in its low-pressure orthorhombic structure (a) in three dimensions, and (b) projected along the plane $\perp\mathbf{a}$. Views of UTe$_2$ in its high-pressure tetragonal structure (c) in three dimensions, and (d) projected along the plane $\perp\mathbf{a}$. In (a,b), the two-legs U-atoms ladders and their nearest Te-atoms neighbors, and the plane perpendicular to $n$, of Miller indices $(0,1,1)$ are emphasized by different sets of colors. The same colors are kept in (c,d) to visualize the effect of the pressure-induced structural transition on the atoms. U atoms with magenta color and Te atoms with blue color form a unit cell in the tetragonal structure.}
\end{figure*}

\subsection{Magnetic and superconducting properties in the orthorhombic phase}
\label{SectionIVA}

Our study of UTe$_2$ in a magnetic field tilted by 30~$^\circ$ from $\mathbf{b}$ towards $\mathbf{c}$ confirms that the metamagnetic field $H_m$ collapses in the vicinity of the critical pressure $p_c$, which was first reported in [\onlinecite{Ran2021}] (and in [\onlinecite{Valiska2021}] but with a sample misorientation). In addition, we find here the Fermi-liquid coefficient $A$ is maximum at $H_m$, and that its maximal value is enhanced near $p_c$, where $H_m$ collapses. Knowing that the coefficient $A$ is presumably driven by the magnetic fluctuations in the systems, its variation in the ($p,H$) plane indicates the presence of stronger magnetic fluctuations at $H_m$ and near $p_c$. The magnetic fluctuations have been proposed to play a crucial role for the development of superconductivity in UTe$_2$, as in other heavy-fermion compounds. Their enhancement near $p_c$ may be responsible for the development of the pressure-induced superconducting phase SC2, while their enhancement near $H_m$ may be responsible for the development of the field-induced superconducting phases SC2 (suspected, but not definitively proved, to be the same phase than that induced under pressure [\onlinecite{Lin2020},\onlinecite{Kinjo2023}]) and SC-PPM near $H_m$ [\onlinecite{Ran2019b,Knafo2021}]. In [\onlinecite{Ran2021}], a study combining temperatures down to 400~mK, steady fields up to 45 T and pressures up to 1.54~GPa permitted to show an extension of the phase SC-PPM under pressure, with critical superconducting temperatures of less than 1~K. Due to the temperatures $T\geq1.4$~K investigated here, and perhaps to a non optimum tilting of the field direction, we could not observed the superconducting phase SC-PPM in our experimental data. At temperatures below $T_N = 3.7 $~K, we could observe the signature of a moment reorientation inside the antiferromagnetic phase. At $p = 2.4 $~GPa and $T=1.4$~K, a moment reorientation occurs at $\mu_0H_r=2.5$~T and antiferromagnetism is destroyed beyond $\mu_0H_c=13.5$~T. The variations of $\rho_{xx}$ at $H_r$ and $H_c$ resemble that observed at $H_{sf}$ and $H_c$ in the prototypical Heisenberg antiferromagnet YbNiSi$_3$ [\onlinecite{Grube2006}]. In YbNiSi$_3$, as in other weakly-anisotropic antiferromagnets, a spin-flop transition is induced at a magnetic $H_{sf}$ applied parallel to the antiferromagnetic-moment directions [\onlinecite{Knafo2021c}]. In UTe$_2$, the magnetic anisotropy is strongly reduced at the critical pressure, as indicated by magnetic susceptibility measurements [\onlinecite{Li2021}], but moment reorientation processes have also been observed under a magnetic field applied along $\mathbf{c}$. Neutron diffraction experiments have also shown that the antiferromagnetic order induced under pressure is associated with an incommensurate propagation vector $\mathbf{k_m}=(0.07,0.67,0)$ [\onlinecite{Knafo2023}]. A component $\mu_m^\perp=0.3$~$\mu_b$/U of the antiferromagnetic moments perpendicular to $\mathbf{b}$ was extracted, indicating that the full antiferromagnetic moment have an amplitude $\mu_m\geq0.3$~$\mu_b$/U, but the magnetic structure could not be resolved. Assuming a complex magnetic structure (for instance a helical incommensurate structure in which the moment direction would rotate elliptically in space), subtle effects could be induced in a magnetic field (for instance a selection of $\mathbf{k}$-domains or a field-induced change of magnetic wavevector $\mathbf{k}$; see CeRhIn$_5$ [\onlinecite{Raymond2007}]). In the future, efforts are needed to elucidate the antiferromagnetic structure of UTe$_2$ and its modifications by a magnetic field, as that observed indirectly at $H_r$ by electrical resistivity here.

\subsection{Field direction in the tetragonal phase}
\label{SectionIVB}

Honda \textit{et al} proposed three scenarios to describe the structural transition of UTe$_2$ under pressure [\onlinecite{Honda2023}]. In each scenario a given group of atoms in the orthorhombic structure was identified as leading to an elementary cell of the tetragonal structure (see Supplemental Figure S13 in the Supplemental Material [\onlinecite{SM}]). One of the three scenarios, the first one proposed by Honda \textit{et al}, seems more likely than the two others. This scenario implies the smallest translations of the atoms and the smallest tilts of the atomic bonds at the structural transition, which contrasts with the more complex sets of atomic rearrangements needed for the second and third scenarios. In the following, we describe the structural transition with a given set of atomic displacements compatible with the first scenario. This allows us to determine the direction of the magnetic field applied here relatively to the high-pressure tetragonal-phase main directions (see next paragraph), but also to discuss how a uniaxial pressure may affect the structural transition (see Section \ref{SectionIVC}).

\begin{figure}[t]
\includegraphics[width=0.85\columnwidth,trim={1cm 11.5cm 18cm 0cm},clip]{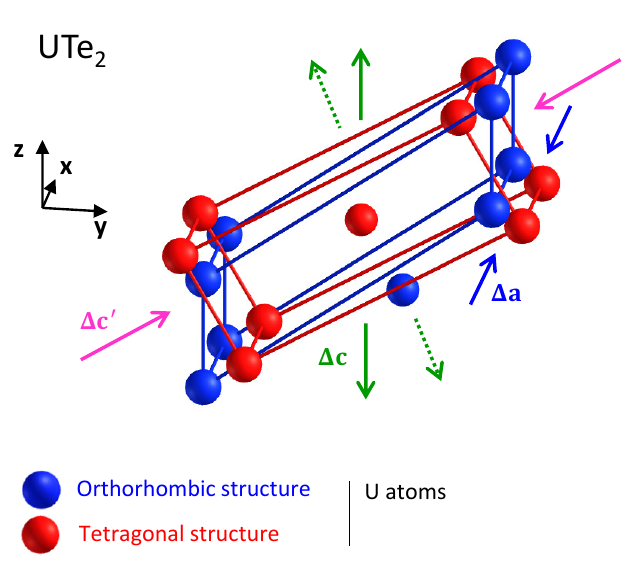}
\caption{\label{Fig11} Comparison of the positions of a group of atoms forming a unit cell in the high-pressure tetragonal structure and their positions in the low-pressure orthorhombic structure, emphasizing the lattice distortions (indicated schematically by arrows) induced at the structural transition.}
\end{figure}

Figure \ref{Fig10} presents three-dimensional views and projections along the $(\mathbf{b},\mathbf{c})$ plane (directions defined in the orthorhombic structure) of UTe$_2$ atoms in their orthorhombic phase at $p=1$~bar [Insets (a,b)] and in their tetragonal phase at $p>3-8$~GPa [Insets (c,d)]. The low-pressure orthorhombic structure ($Immm$) was drawn assuming the lattice parameters $a=4.16$~\AA, $b=6.12$~\AA, and $c=13.96$~\AA, and the parameter $z=0.13544$ (position of the U atom in the cell) obtained at $p=1$~bar and $T=300$~K [\onlinecite{Ikeda2006}], and the high-pressure tetragonal structure ($I4/mmm$) was drawn assuming the lattice parameters $a'=3.89$~\AA~and $c'=9.80$~\AA~obtained at $p=4$~GPa and $T=300$~K [\onlinecite{Honda2023}] (see unit cells of the two structures in Figure \ref{Fig1}). U atoms with magenta color and Te atoms with blue color form a unit cell in the high-pressure tetragonal phase and are also identified in the low-pressure orthorhombic phase, i.e., prior to the structural transition. In the Supplemental Material [\onlinecite{SM}], we detail how the structural transition can be artificially decomposed into five elementary steps corresponding to a given set of atomic displacements or lattice distortions. As shown in Figure \ref{Fig10}, the planes of Miller indices (0 1 1) and normal to the direction $\mathbf{n}$ in the low-pressure orthorhombic structure [\onlinecite{Knafo2021}], are transformed into the basal planes $\perp\mathbf{c}'$ in the high-pressure tetragonal structure. In the present study, the magnetic field $\mathbf{H}$ was applied in a direction tilted by 30~$^\circ$ from $\mathbf{b}$ towards $\mathbf{c}$ in the low-pressure orthorhombic phase. This direction is very close to the direction $\mathbf{n}$, which is tilted by 23.7~$^\circ$ from $\mathbf{b}$ towards $\mathbf{c}$, indicating that $\mathbf{H}$ was applied near to the direction $\mathbf{c}'$ in the high-pressure tetragonal phase.

In addition to their role for the structural transition, the planes of Miller indices (0 1 1) are an important feature of the orthorhombic structure. It was noticed that they are cleaving planes of the crystal and a possible relation with the stabilization of the field-induced superconducting phase SC-PPM in a magnetic field $\mathbf{H}\simeq\parallel\mathbf{n}$ was emphasized [\onlinecite{Knafo2021}]. It is also clear from Figures \ref{Fig10}(a,b) that, in the low-pressure orthorhombic phase, Te atoms are already almost lying within planes $\perp\mathbf{n}$. Assuming a larger electronegativity of Te atoms, in comparison with that of U atoms, these Te planes may constitute negative-charge reservoirs. A question is whether they could play a role for the stabilization of the different superconducting phases of UTe$_2$ in its orthorhombic structure, for instance the phase SC-PPM induced in magnetic fields $\mathbf{H}\simeq\parallel\mathbf{n}$ beyond $H_m$. Another question is how the valence change observed at the structural transition [\onlinecite{Wilhelm2023},\onlinecite{Deng2024}], which could be related with a modification of the charges on Te and U atoms, may be involved in the drastic change of the magnetic and superconducting properties in the tetragonal phase.

\subsection{Effects of a uniaxial pressure}
\label{SectionIVC}

Different critical pressures $p^*$, ranging from 3 to 8~GPa, of the structural transition were determined from the study of powdered samples and single crystals of UTe$_2$ (see here and in [\onlinecite{Valiska2021},\onlinecite{Valiska2021SI},\onlinecite{Huston2022},\onlinecite{Honda2023}]). In [\onlinecite{Huston2022}] (x-ray diffraction), non-hydrostatic conditions of pressure on powdered samples were identified as driving to the increase of $p^*$, from 5~GPa under hydrostatic pressure up to 8~GPa under non-hydrostatic pressure. In [\onlinecite{Honda2023}] (x-ray diffraction and electrical resistivity), a smaller critical pressure $p^*\simeq3.5-4$~GPa was found for single crystals of main faces with Miller indices (1 0 0) and (1 1 0). In the same work, a critical pressure $p^*\simeq5$~GPa was determined for powder crystals, for which the orthorhombic and tetragonal domains were found to coexist under pressures from 5 to 7~GPa. In the studies presented here and in [\onlinecite{Valiska2021},\onlinecite{Valiska2021SI}] (electrical resistivity), signatures of the tetragonal phase were observed under pressures $p>3.1$~GPa, indicating a critical pressure $p^*\lesssim3$~GPa, for $\mathbf{H}$ tilted by $\simeq25-30~^\circ$ from $\mathbf{b}$ to $\mathbf{c}$ (see Supplemental Material [\onlinecite{SM}]). Oppositely, the signatures of the high-pressure tetragonal phase were not observed under pressures $p\leq 4$~GPa, indicating a critical pressure $p^*\gtrsim4$~GPa, for $\mathbf{H}$ applied along $\mathbf{c}$ [\onlinecite{Valiska2021},\onlinecite{Valiska2021SI}].

Here, we attempt to discuss the effects on single crystals from a non-hydrostatic pressure in the cells used here and [\onlinecite{Valiska2021},\onlinecite{Valiska2021SI}].
As shown schematically in Figure \ref{Fig11}, the stabilization of the high-pressure tetragonal phase is associated with a contraction of the lattice along the direction $\mathbf{c}'$ of the tetragonal structure (or almost equivalently along the direction $\mathbf{n}$ of the orthorhombic structure) and with an expansion along the directions $\mathbf{a}$ and $\mathbf{c}$ of the orthorhombic structure. Uniaxial pressures applied parallely or perpendicularly to a direction close to $\mathbf{n}$ may therefore favor or unfavor, respectively, the establishment of a tetragonal phase in comparison the effect of a purely-hydrostatic pressure [\onlinecite{Huston2022,Honda2023,Deng2024}]. The pressure cell used here and in [\onlinecite{Valiska2021},\onlinecite{Valiska2021SI}] was setup with its vertical axis parallel to the magnetic field, indicating that uniaxial pressures may have been generated along a direction parallel to the field. This may explain why the signatures of the structural transition were observed above moderate pressures of $\simeq3$~GPa in a field tilted by $\simeq30~^\circ$ from $\mathbf{b}$ towards $\mathbf{c}$ in [\onlinecite{Valiska2021},\onlinecite{Valiska2021SI}] and here, but not observed in fields $\mathbf{H}\parallel\mathbf{c}$ under pressures up to 4~GPa in [\onlinecite{Valiska2021},\onlinecite{Valiska2021SI}].

We note that the critical pressure $p^*$ of the structural transition also depends on the temperature [\onlinecite{Honda2023}]. Different values of $p^*$ are therefore expected, depending if the pressure was changed at low temperature (as in [\onlinecite{Honda2023},\onlinecite{Braithwaite2019}] or at room temperature (as here and in [\onlinecite{Valiska2021},\onlinecite{Huston2022},\onlinecite{Deng2024}].

\subsection{Magnetic order in the tetragonal phase ?}
\label{SectionIVD}

Different elements support that magnetic order may be stabilized below the transition temperature $T_x$ in the tetragonal phase of UTe$_2$ (see also discussion in [\onlinecite{Honda2023}]). The electrical resistivity presents a similar variation at $T_x$ than that observed at the transition temperature of other uranium compounds where high-temperature magnetic order was evidenced. Examples are the ferromagnets UTe, USe and US associated with the Curie temperatures $T_C$ $\simeq100$~K, 160~K, and 177~K, respectively [\onlinecite{Schoenes1986}], and the antiferromagnets UAs$_2$ and USb$_2$ associated with the N\'{e}el temperatures $T_N$ = 270 K and 202 K, respectively [\onlinecite{Wisniewski2000},\onlinecite{Wawryk2006}]. Step-like variations of the low-temperature electrical resistivity of UTe$_2$ were also observed at the magnetic fields $H_{x1}$ and $H_{x2}$. They transform into kinks in the electrical resistivity at temperatures $T\lesssim T_x = 235$~K and disappear at temperatures larger than $T_x$. The transitions at $H_{x1}$ and $H_{x2}$ are therefore a property of the electronic state stabilized below the transition temperature $T_x$. Assuming that $T_x$ is the temperature of a magnetic phase transition, we identify the anomalies at $H_{x1}$ and $H_{x2}$ as the signatures of metamagnetic transitions, i.e., first-order transitions associated with a sudden increase of the magnetization induced by a reorientation of the magnetic moments. In the case of a ferromagnet with a strong magnetic anisotropy, metamagnetism can be induced by a magnetic field applied perpendicular to the ferromagnetic moments, as observed for instance in URhGe [\onlinecite{Levy2005}]. However, metamagnetism is generally observed in antiferromagnets, either with a small magnetic anisotropy ('spin-flop' transition) or with a strong magnetic anisotropy, when the magnetic field is applied parallel to the antiferromagnetic-moments direction [\onlinecite{Knafo2021c}]. Step-like variations in the magnetization and magnetostriction were for instance observed in the antiferromagnet USb$_2$ at temperatures smaller than $T_N$, indicating a first-order metamagnetic transition [\onlinecite{Stillwell2017}]. In UTe$_2$, antiferromagnetic moments may be fully aligned, or partly-tilted, along or near to the direction $\mathbf{c}'$ of the tetragonal structure. Alternatively, we cannot exclude that the transitions at $T_x$ and $H_x$ may be driven by a valence transition, as observed for instance in YbInCu$_4$ [\onlinecite{Matsuda2007}]. We also note that the two transitions fields $H_{x1}$ and $H_{x2}$ observed here, which share a similar temperature evolution up to $T_x$, may be induced by two crystalline domains formed under pressure and aligned differently relatively to the magnetic field. A single transition field $H_x$ was observed in the study made in [\onlinecite{Valiska2021},\onlinecite{Valiska2021SI}] suggesting then a dominant domain. The possibility of crystalline domains in the present study may be related with the observation by x-ray diffraction of two domains in the high-pressure tetragonal phase [\onlinecite{Honda2023}]. In the future, challenges will be to characterize the nature of the electronic state, possibly of magnetic origin as suspected here, which is established below the transition temperature $T_x$, and to study if the formation of domains may be related with the doubling of the metamagnetic transition, or if multiple metamagnetic transitions could be an intrinsic high-field property of the system.

\section*{Acknowledgements}

We acknowledge useful discussion with Fuminori Honda, and financial support from the French National Research Agency collaborative research project FRESCO No. ANR-20-CE30-0020, from the Cross-Disciplinary Program on Instrumentation and Detection of CEA, the French Alternative Energies and Atomic Energy Commission, and from the JSPS KAKENHI Grants Nos. JP19H00646, 19K03756, JP20H00130, 20H01864, JP20K20889, JP20KK0061, and 21H04987.

\newpage

\onecolumngrid

\setcounter{section}{0}
\setcounter{figure}{0}
\renewcommand\thefigure{S\arabic{figure}}
\renewcommand{\theequation}{S\arabic{equation}}
\renewcommand{\thetable}{S\arabic{table}}
\renewcommand{\thesection}{S\arabic{section}}
\renewcommand{\bibnumfmt}[1]{[S#1]}
\renewcommand{\citenumfont}[1]{S#1}
\setcounter{figure}{0}
\renewcommand{\thesubsection}{S\arabic{subsection}}

\vspace{15cm}
\begin{center}
\large {\textbf {Supplemental Material:\\ Metamagnetism in the high-pressure tetragonal phase of UTe$_2$}}
\end{center}
\vspace{1cm}

\setcounter{page}{1}

In this Supplemental Material, we present complementary graphs about raw data obtained in our electrical-resistivity study of UT$_2$ under pulsed magnetic fields and the analysis made to extract the phase diagrams and the Fermi-liquid coefficient $A$. Atomic displacements at the pressure-induced structural transition are discussed and a comparison of the results obtained here and those from a previous study published in [Vali{\v{s}}ka \textit{et al.}, Phys. Rev. B \textbf{104}, 214507 (2021)] is made.

\newpage

\section{Supplementary electrical-resistivity data}
\label{raw_data_SM}

\begin{figure}[t]
\includegraphics[width=1\columnwidth]{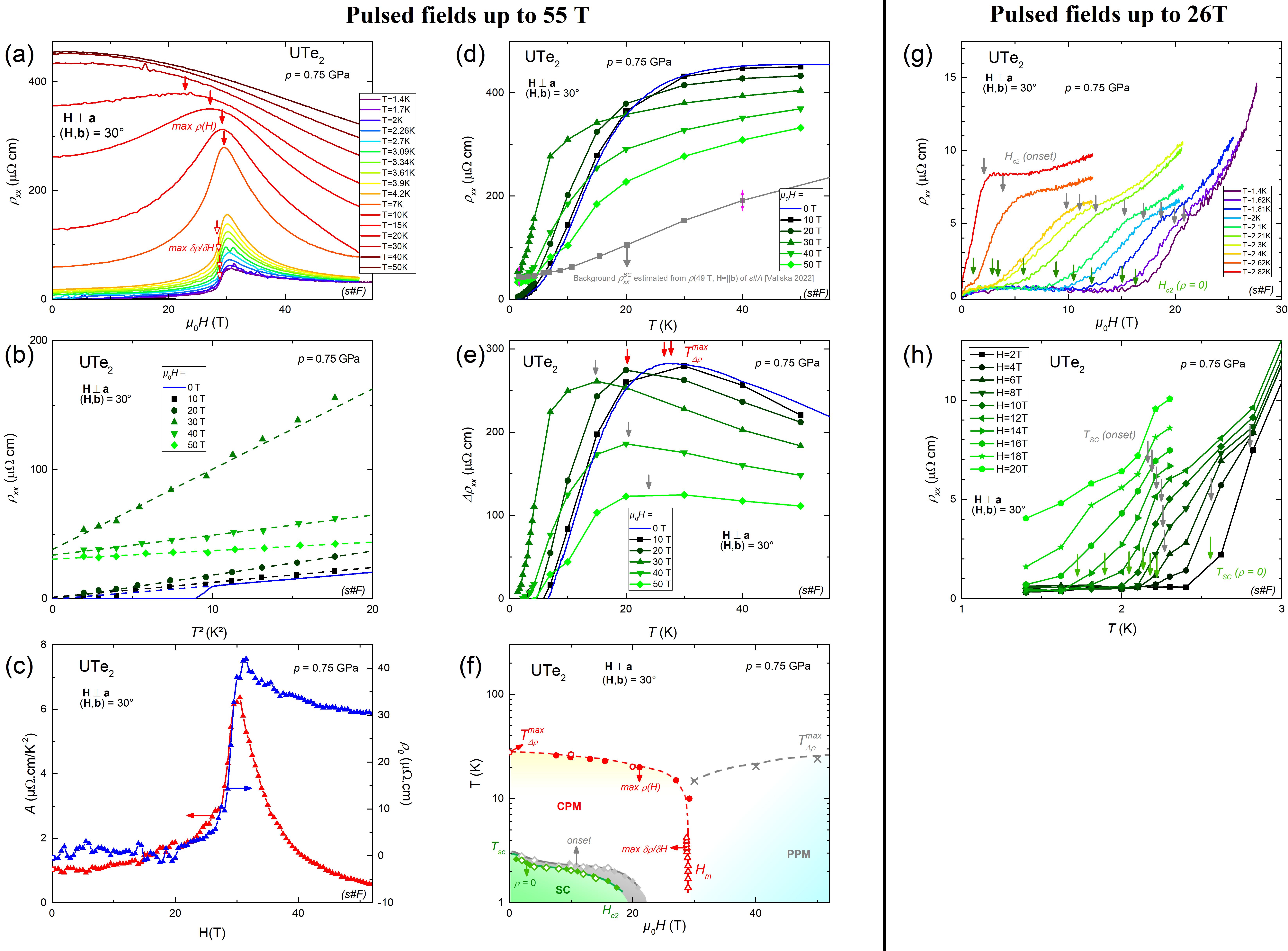}
\caption{\label{FigS1} Details about electrical resistivity versus different magnetic field and temperature data, the resulting magnetic-field-temperature phase diagram, and the Fermi-liquid $\rho_{xx}=\rho_0+A T^2$ fits to the data at the pressure $p=0.75$~GPa.}
\end{figure}

\begin{figure}[t]
\includegraphics[width=1\columnwidth]{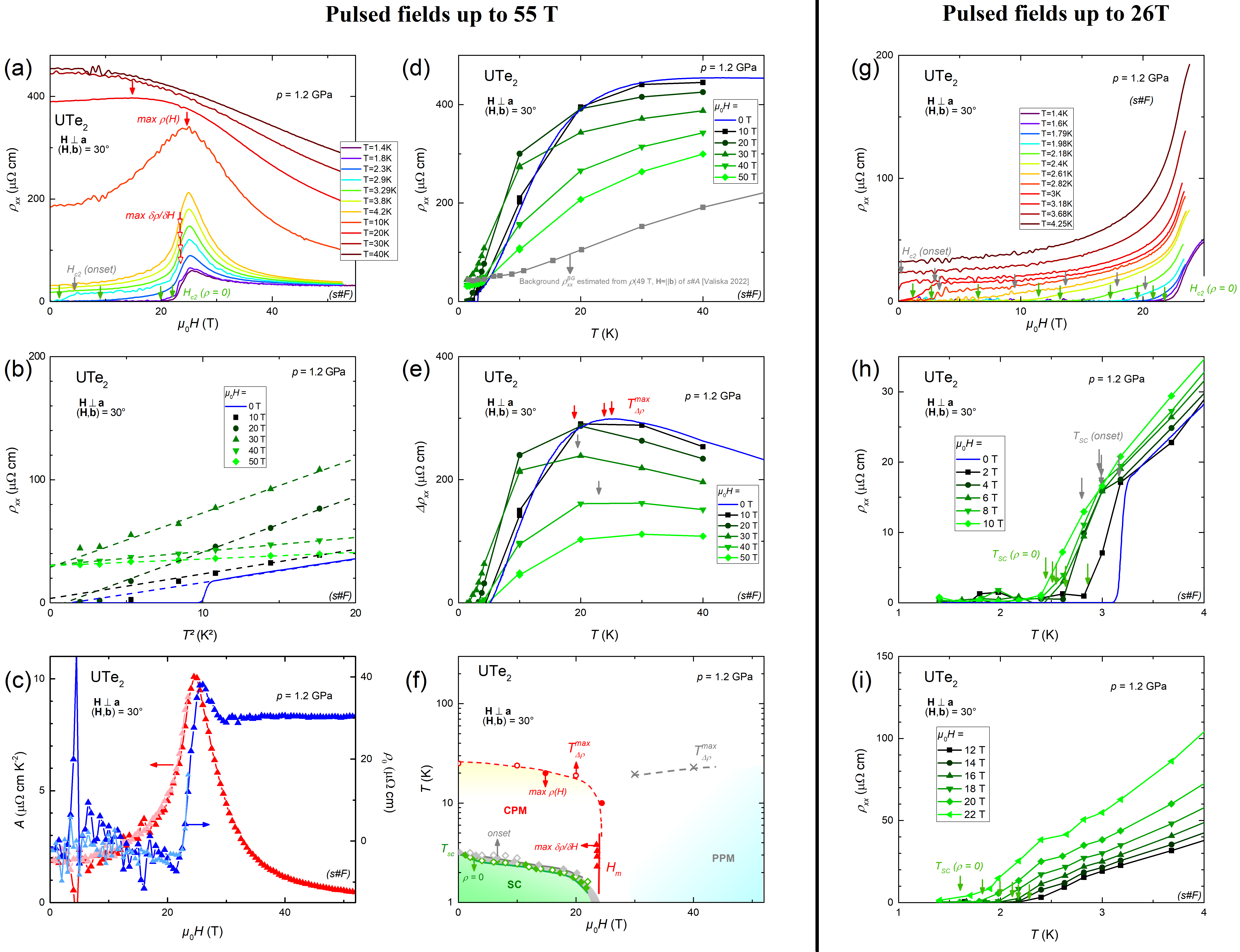}
\caption{\label{FigS2} Details about electrical resistivity versus different magnetic field and temperature data, the resulting magnetic-field-temperature phase diagram, and the Fermi-liquid $\rho_{xx}=\rho_0+A T^2$ fits to the data at the pressure $p=1.2$~GPa.}
\end{figure}

\begin{figure}[t]
\includegraphics[width=1\columnwidth]{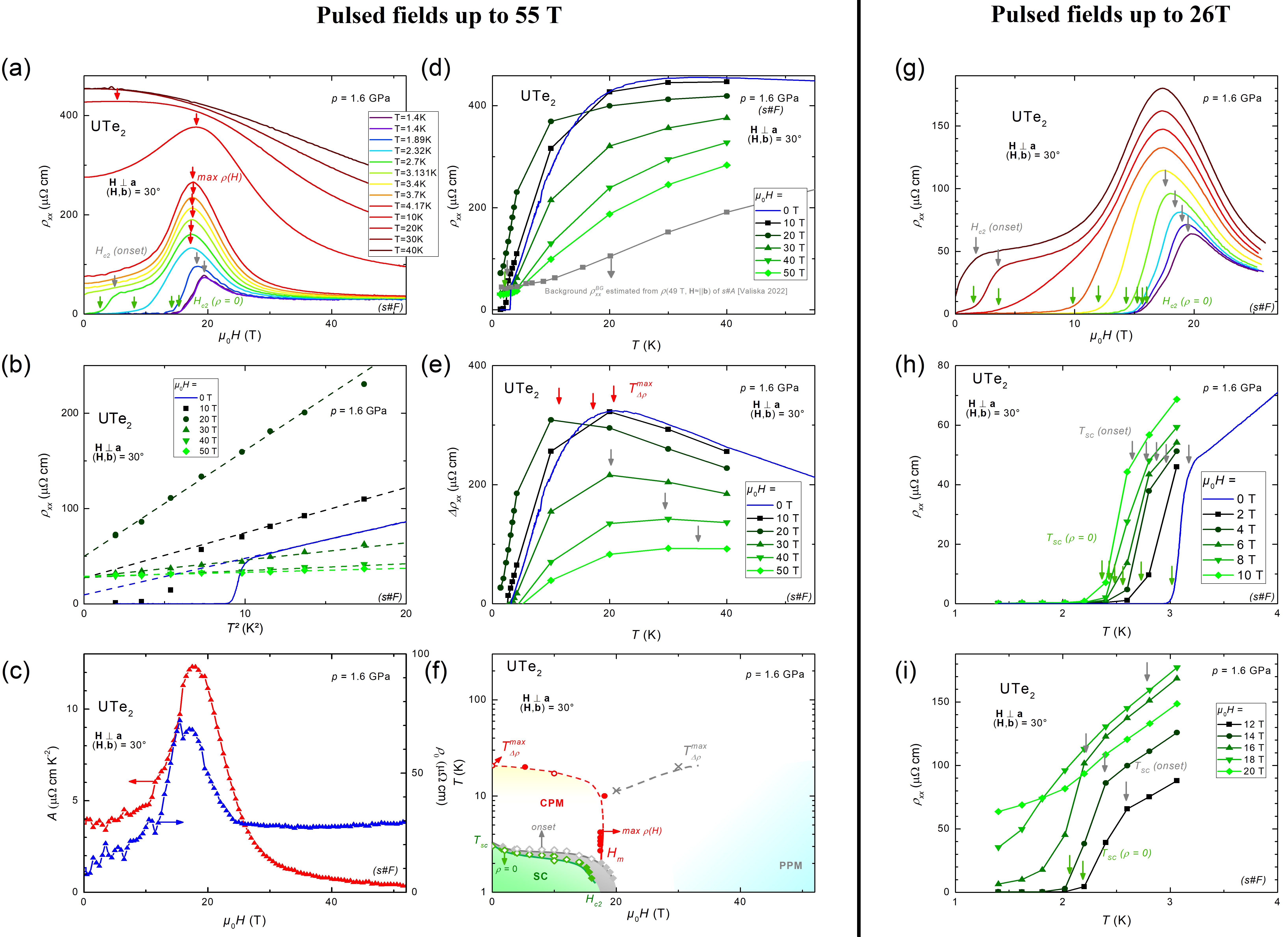}
\caption{\label{FigS3} Details about electrical resistivity versus different magnetic field and temperature data, the resulting magnetic-field-temperature phase diagram, and the Fermi-liquid $\rho_{xx}=\rho_0+A T^2$ fits to the data at the pressure $p=1.6$~GPa.}
\end{figure}

\begin{figure}[t]
\includegraphics[width=1\columnwidth]{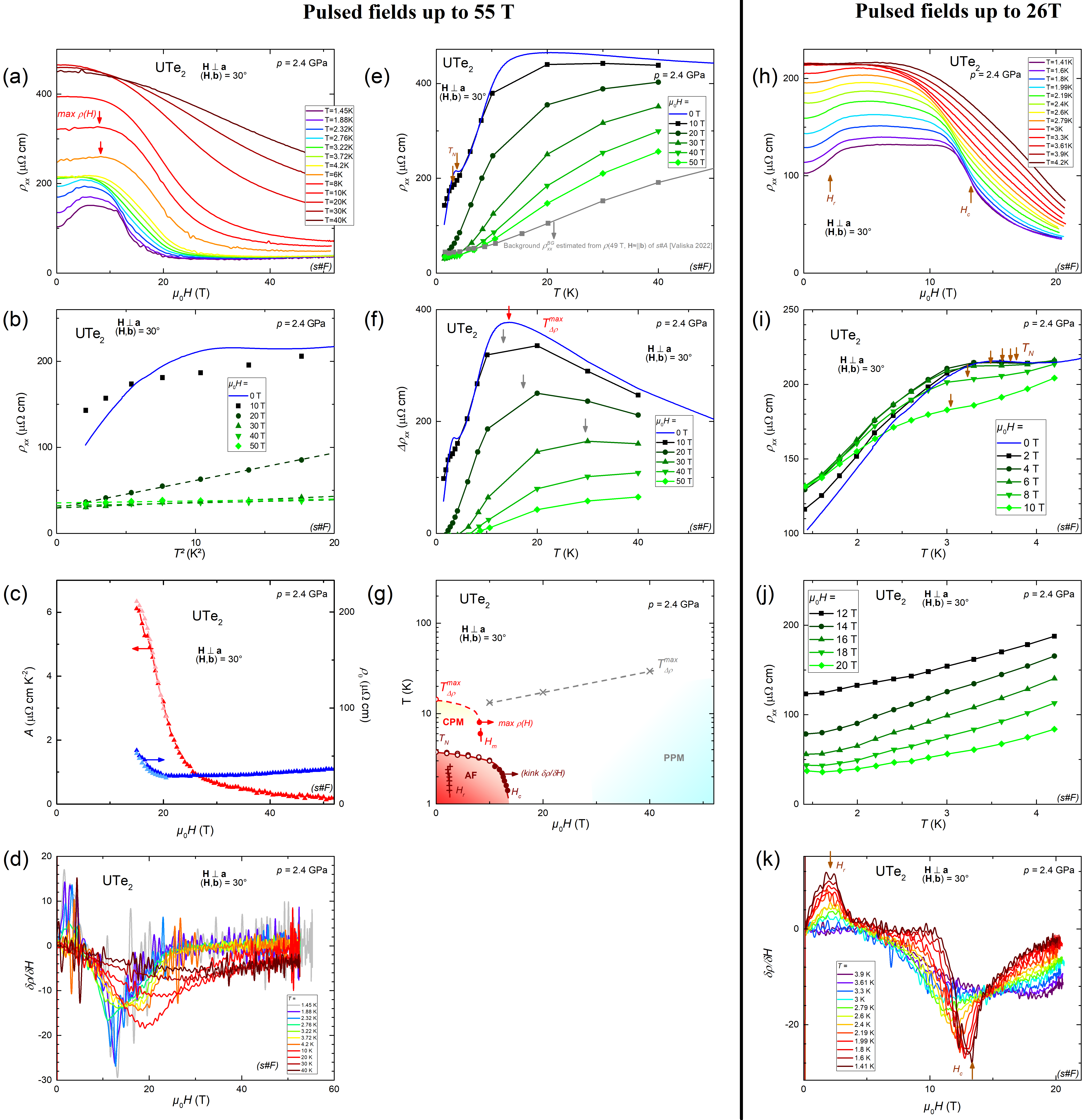}
\caption{\label{FigS4} Details about electrical resistivity versus different magnetic field and temperature data, the resulting magnetic-field-temperature phase diagram, and the Fermi-liquid $\rho_{xx}=\rho_0+A T^2$ fits to the data at the pressure $p=2.4$~GPa.}
\end{figure}

\begin{figure}[t]
\includegraphics[width=1\columnwidth]{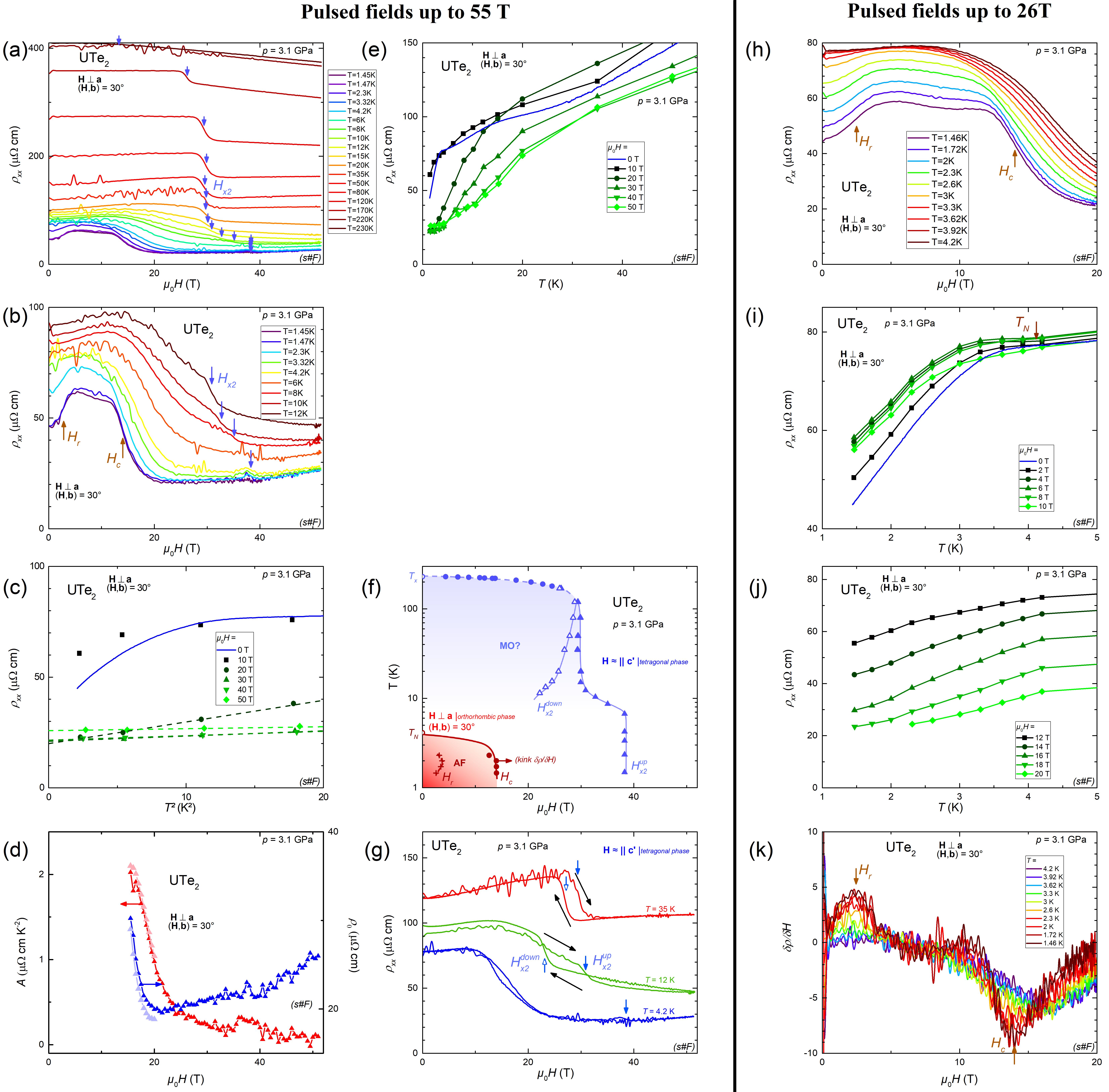}
\caption{\label{FigS5} Details about electrical resistivity versus different magnetic field and temperature data, the resulting magnetic-field-temperature phase diagram, and the Fermi-liquid $\rho_{xx}=\rho_0+A T^2$ fits to the data at the pressure $p=3.1$~GPa.}
\end{figure}

\begin{figure}[t]
\includegraphics[width=1\columnwidth]{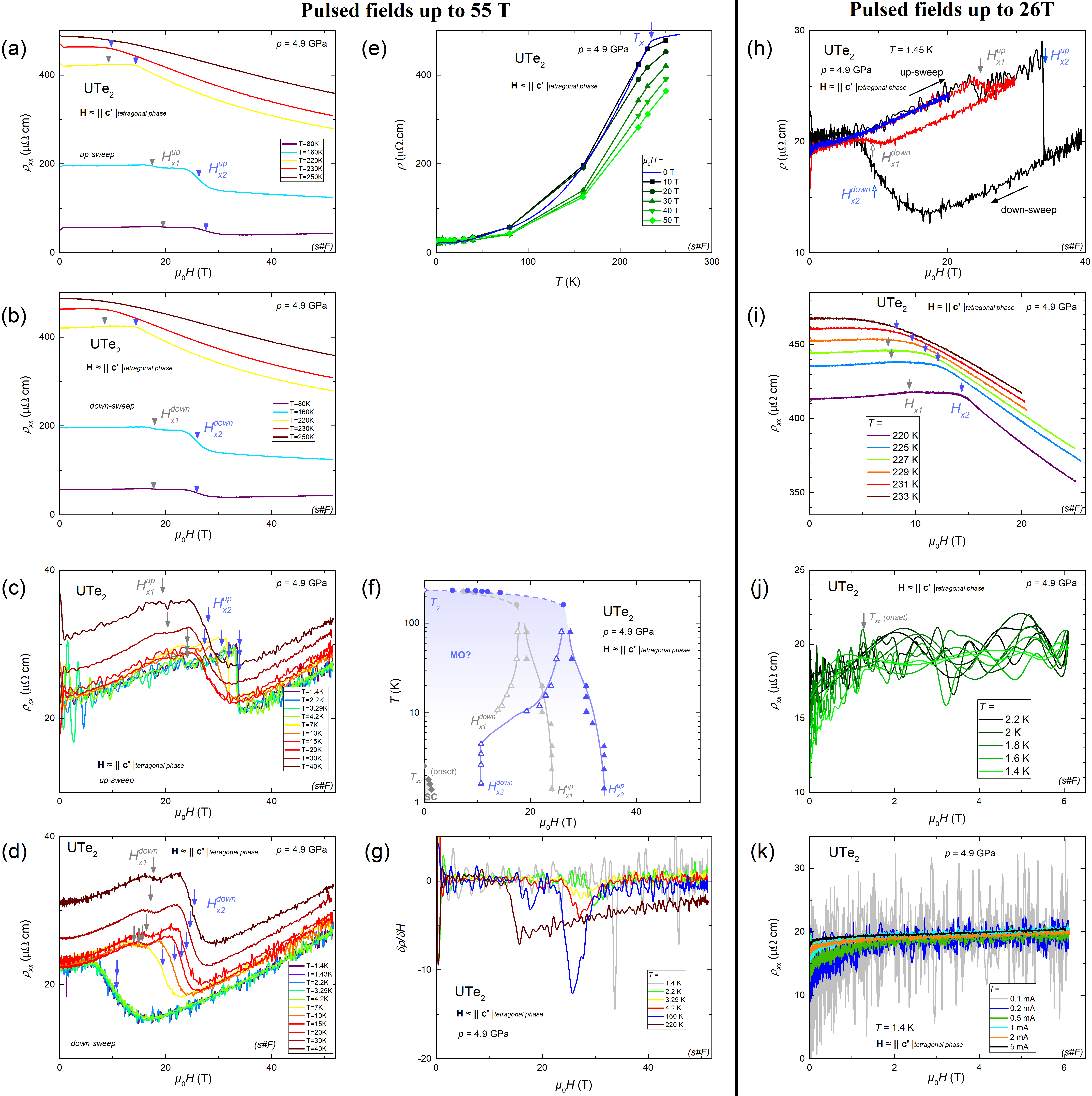}
\caption{\label{FigS6} Details about electrical resistivity versus different magnetic field and temperature data, and the resulting magnetic-field-temperature phase diagram at the pressure $p=4.9$~GPa.}
\end{figure}

\begin{figure}[t]
\includegraphics[width=1\columnwidth]{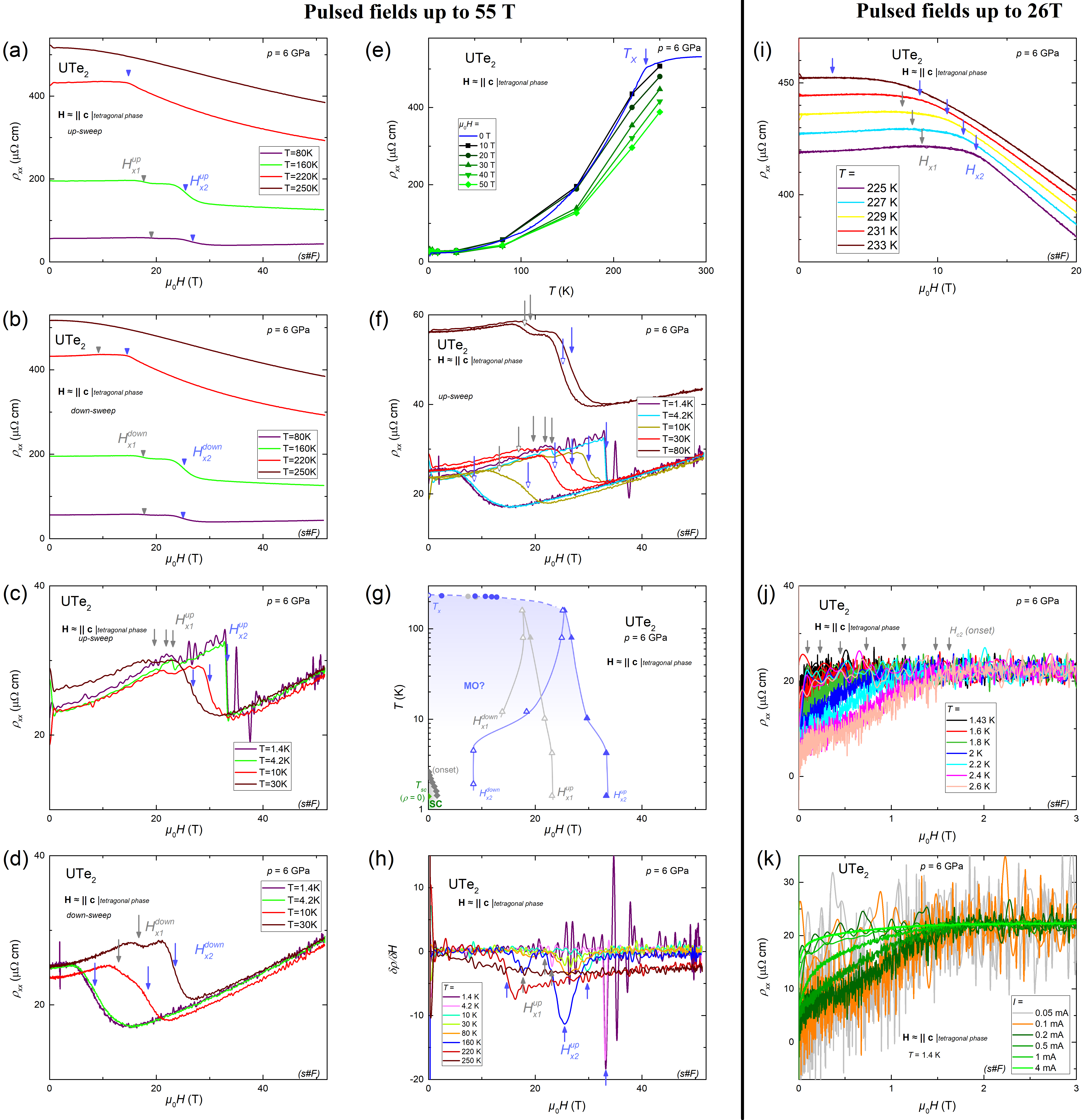}
\caption{\label{FigS7} Details about electrical resistivity versus different magnetic field and temperature data, and the resulting magnetic-field-temperature phase diagram at the pressure $p=6$~GPa.}
\end{figure}

Figures \ref{FigS1}, \ref{FigS2}, \ref{FigS3}, \ref{FigS4}, \ref{FigS5}, \ref{FigS6}, and \ref{FigS7} present details about raw data obtained in this study at the pressures $p=0.75$, 1.2, 1.6, 2.4, 3.1, 4.9, and 6 GPa, respectively, and their analysis which permitted to construct magnetic-field-temperature phase diagrams at all pressures and to extract field dependencies of the Fermi-liquid coefficient $A$ at pressures $p\lesssim3.1$~GPa.

\begin{figure}[t]
\includegraphics[width=0.4\columnwidth]{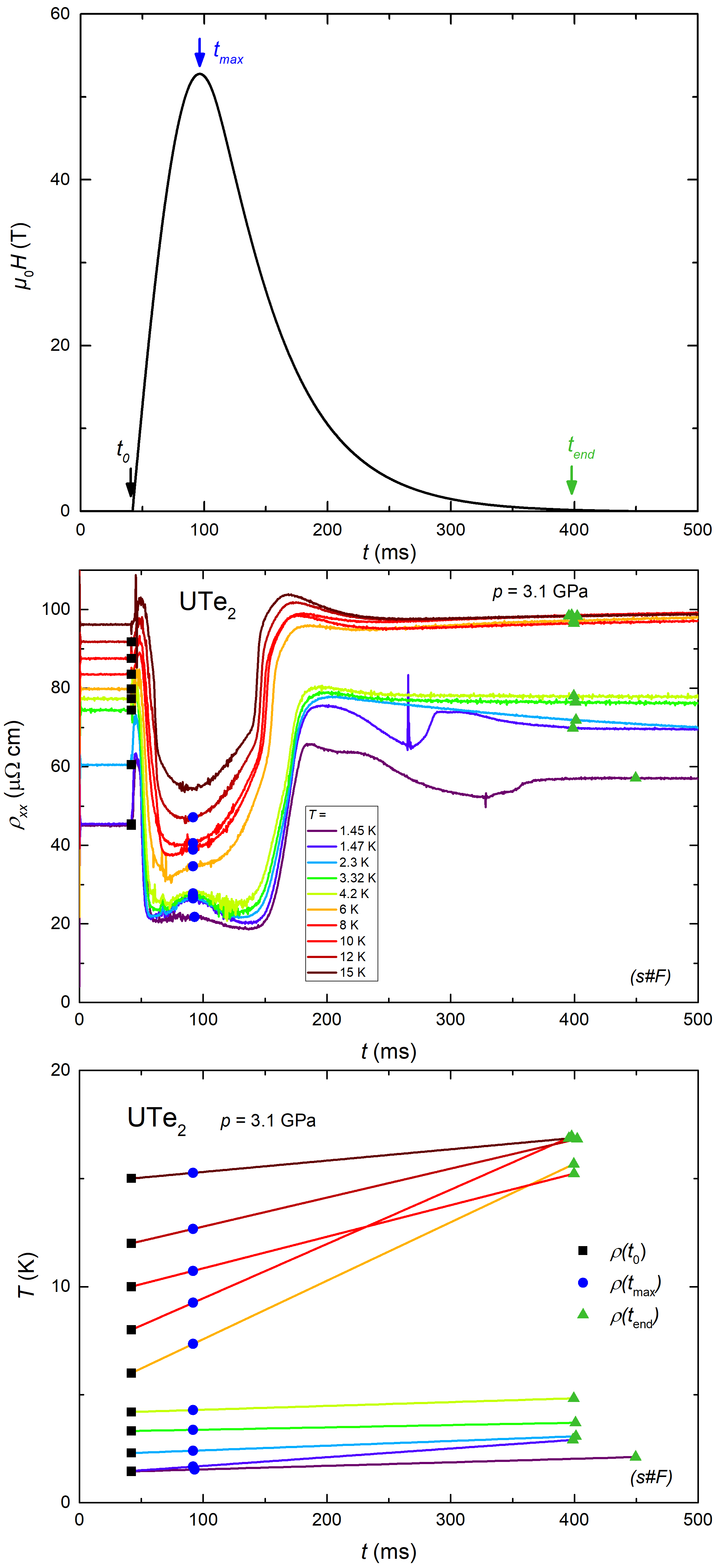}
\caption{\label{FigS8} (a) time profile of a 52-T magnetic-field pulse, (b) time variation of the electrical resistivity measured for in magnetic fields pulsed up to 50-55~T, for different temperatures, and (c) estimation of the heating of the sample, assuming a linear increase of temperature with time between the beginning and end of the pulse. $t_0$, $t_{max}$ and $t_{end}$ denote the time at the beginning, at the maximum field, and at the end of a magnetic-field pulse.}
\end{figure}

While most of the analyzed data were obtained from rising fields, at which sample heating induced by Eddy currents generated in the metallic body of the cell are limited, some of the data were obtained for falling fields, where larger heating are induced for initial temperatures ranging from $>4.2$~K to $\simeq20$~K [\onlinecite{Braithwaite2016S}]. Assuming a linear increase of temperature during the field pulse, which is a rough, but, -we think- acceptable approximation, Figure \ref{FigS8} shows an example where the electrical resistivity at zero field at the end of the pulse was compared with the electrical resistivity at zero field just before the pulse, in order to estimate the increase of temperature of the sample during the pulse. The temperature increases due to Eddy currents in the cell estimated here are in agreement with those made in [\onlinecite{Braithwaite2016S}].

\begin{figure}[t]
\includegraphics[width=1\columnwidth]{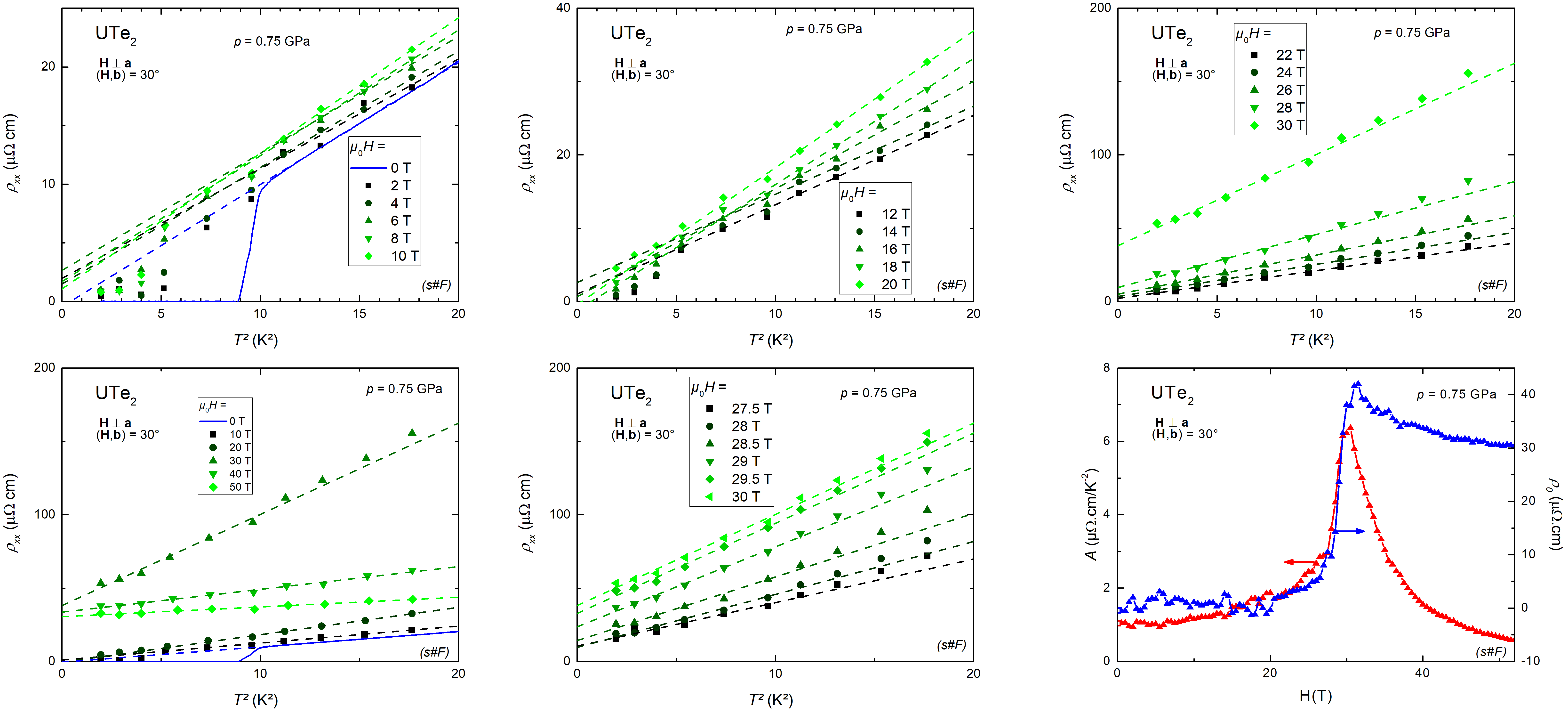}
\caption{\label{FigS9} Details about the Fermi-liquid $\rho_{xx}=\rho_0+A T^2$ fits to the data at the pressure $p=0.75$~GPa.}
\end{figure}

\begin{figure}[t]
\includegraphics[width=1\columnwidth]{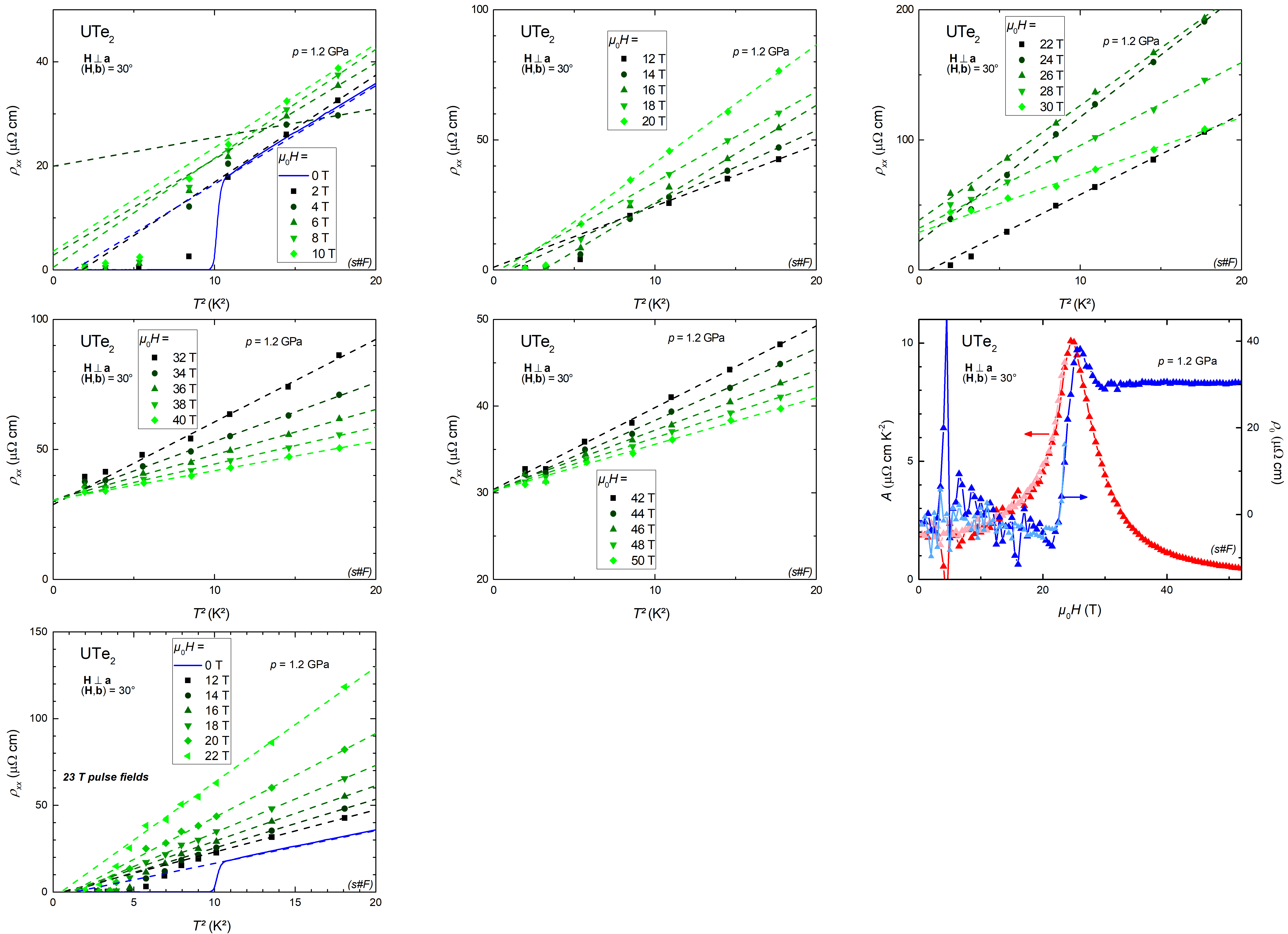}
\caption{\label{FigS10} Details about the Fermi-liquid $\rho_{xx}=\rho_0+A T^2$ fits to the data at the pressure $p=1.2$~GPa.}
\end{figure}

\begin{figure}[t]
\includegraphics[width=1\columnwidth]{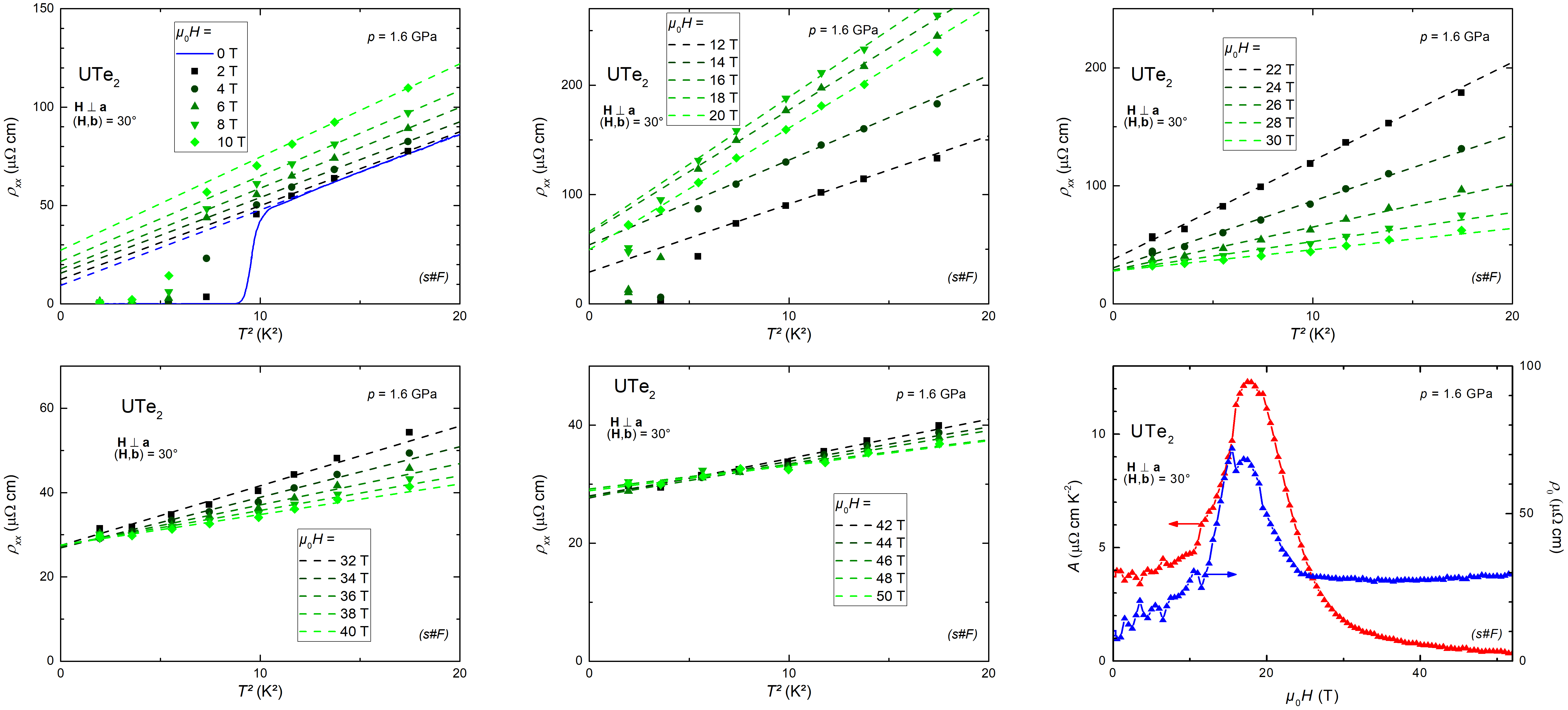}
\caption{\label{FigS11} Details about the Fermi-liquid $\rho_{xx}=\rho_0+A T^2$ fits to the data at the pressure $p=1.6$~GPa.}
\end{figure}

\begin{figure}[t]
\includegraphics[width=1\columnwidth]{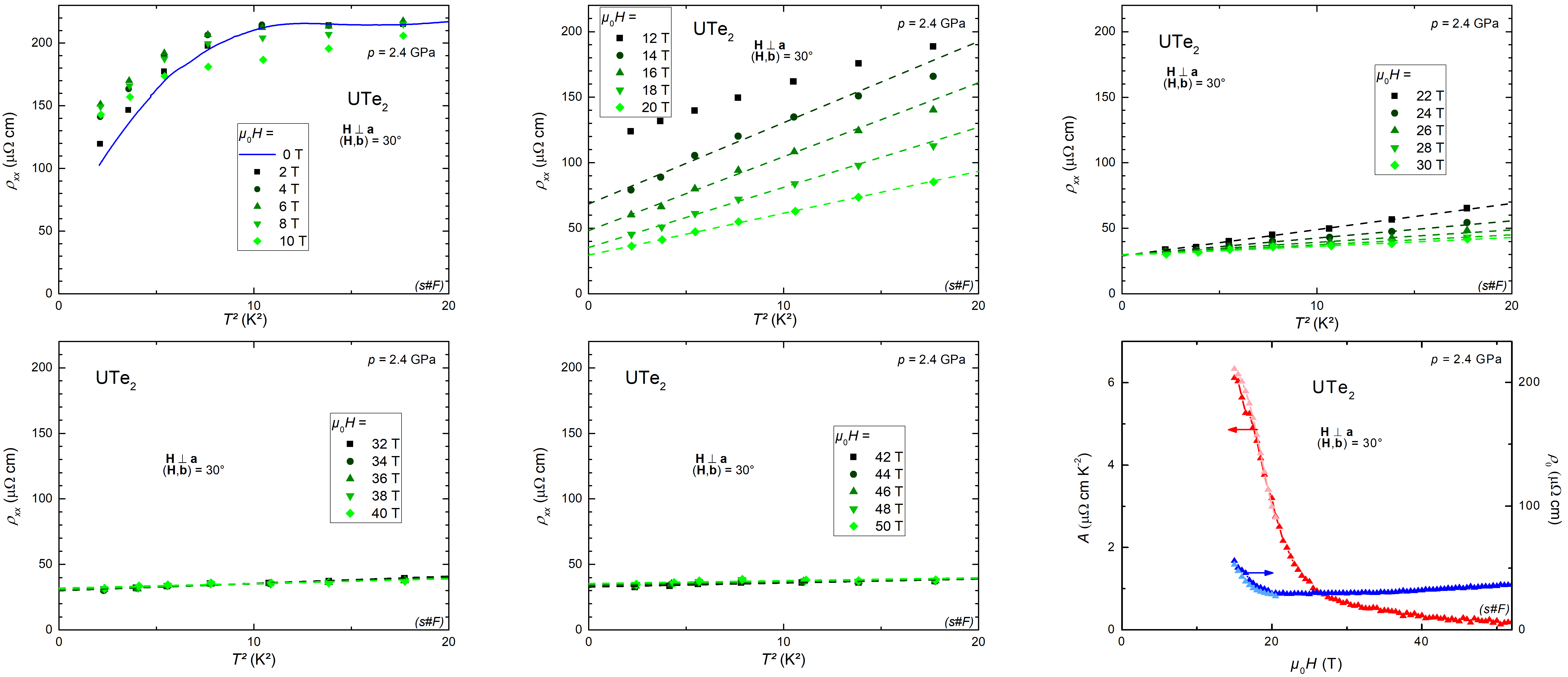}
\caption{\label{FigS12} Details about the Fermi-liquid $\rho_{xx}=\rho_0+A T^2$ fits to the data at the pressure $p=2.4$~GPa.}
\end{figure}

Figures \ref{FigS9}, \ref{FigS10}, \ref{FigS11}, and \ref{FigS12} focus on the fit to the low-temperature electrical resistivity data by a Fermi-liquid formula $\rho_{xx}=\rho_0+AT^2$. Fits were done at temperatures larger than the superconducting transition temperature $T_{sc}$.

\section{Supplementary elements about the structural transition}
\label{structural_transition_SM}

\begin{figure*}[t]
\includegraphics[width=0.5\textwidth,trim={0cm 14cm 9.5cm 0cm},clip]{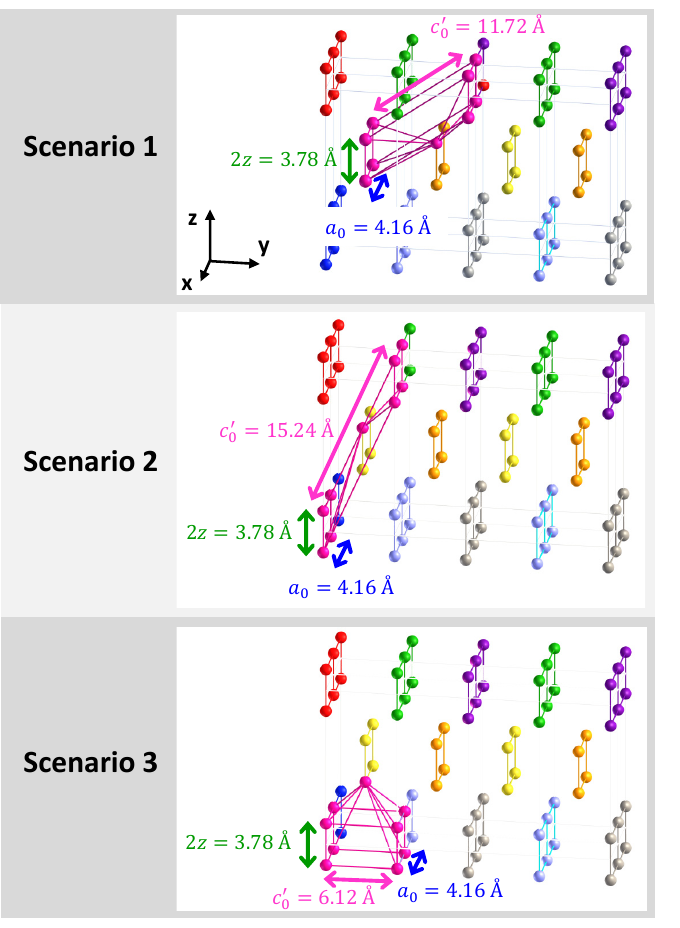}
\caption{\label{FigS13} Three-dimensional views of the lattice of U atoms in UTe$_2$ at ambient pressure extended to several unit cells. The three panels emphasize the three scenarios proposed in [\onlinecite{Honda2023}] to describe the pressure-induced structural transition of UTe$_2$. U atoms, which are indicated with magenta color in the low-pressure orthorhombic phase, form a unit cell in the high-pressure tetragonal phase.}
\end{figure*}

\begin{figure*}[t]
\includegraphics[width=0.7\textwidth,trim={1cm 1cm 3cm 0cm},clip]{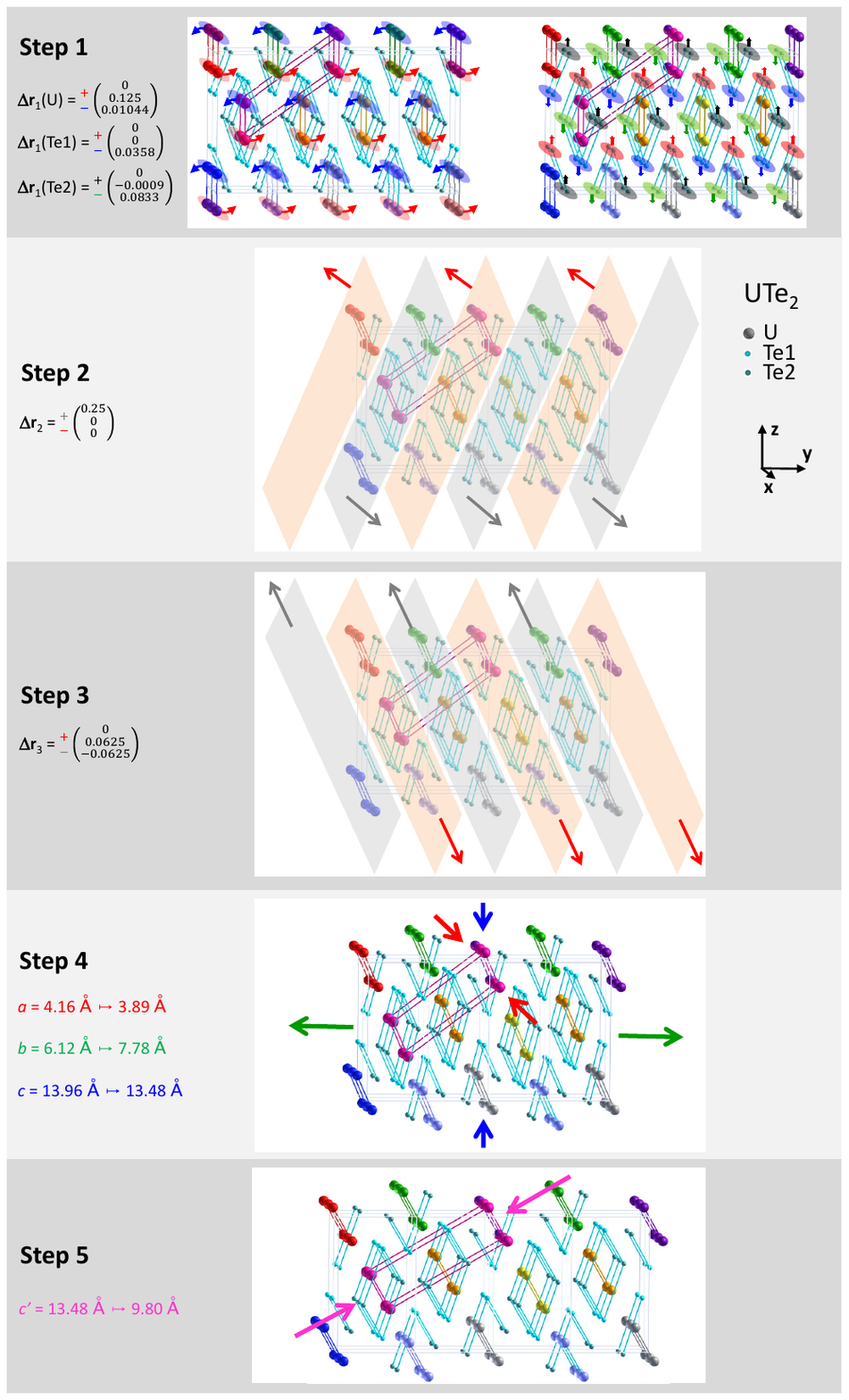}
\caption{\label{FigS14} Decomposition into five artificial steps, consisting in atomic displacements and lattice distortions, of the pressure-induced structural transition of UTe$_2$ assuming the first scenario from [\onlinecite{Honda2023}].}
\end{figure*}

\begin{figure*}[t]
\includegraphics[width=0.7\textwidth,trim={1cm 1cm 4cm 0cm},clip]{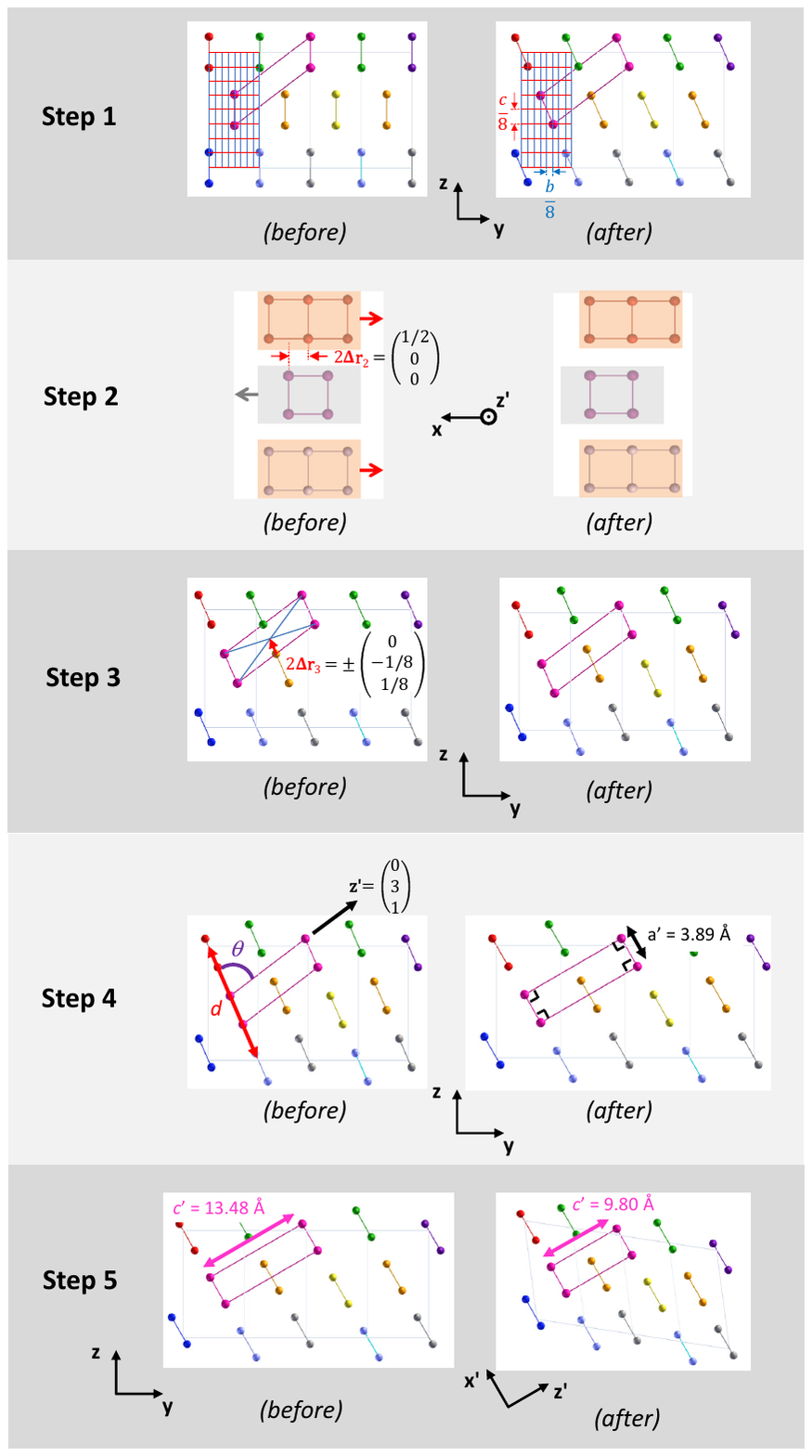}
\caption{\label{FigS15} Details about the five artificial steps of the pressure-induced structural transition of UTe$_2$ assuming the first scenario from [\onlinecite{Honda2023}].}
\end{figure*}

\begin{figure*}[t]
\includegraphics[width=0.9\textwidth,trim={0cm 11cm 0cm 0cm},clip]{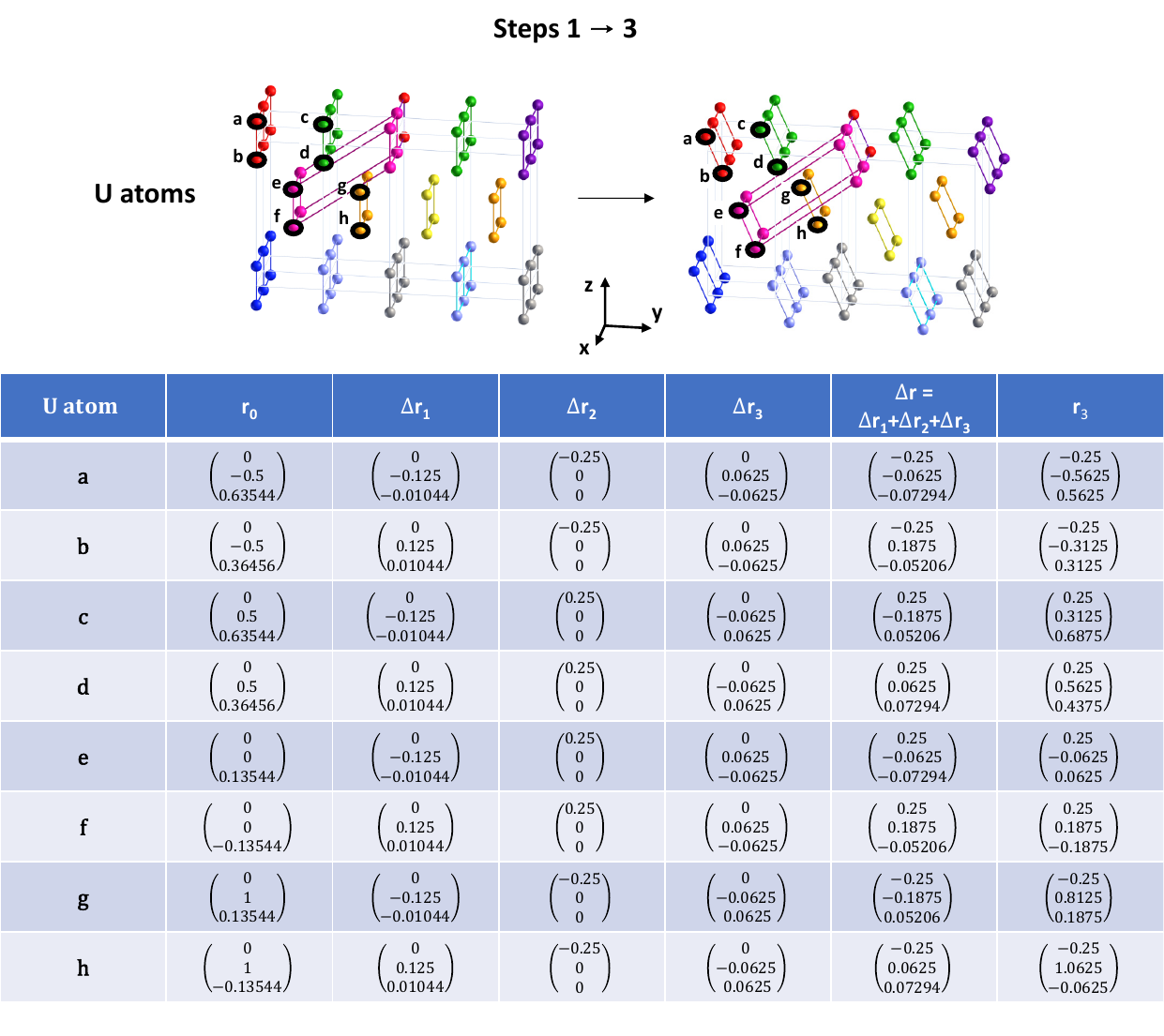}
\caption{\label{FigS16} (top) Identification of U atoms noted \textbf{a} to \textbf{h} and (bottom) Table indicating their displacements in the steps 1 to 3, assuming the first scenario from [\onlinecite{Honda2023}] for the structural transition.}
\end{figure*}

\begin{figure*}[t]
\includegraphics[width=0.9\textwidth,trim={0cm 1cm 0cm 0cm},clip]{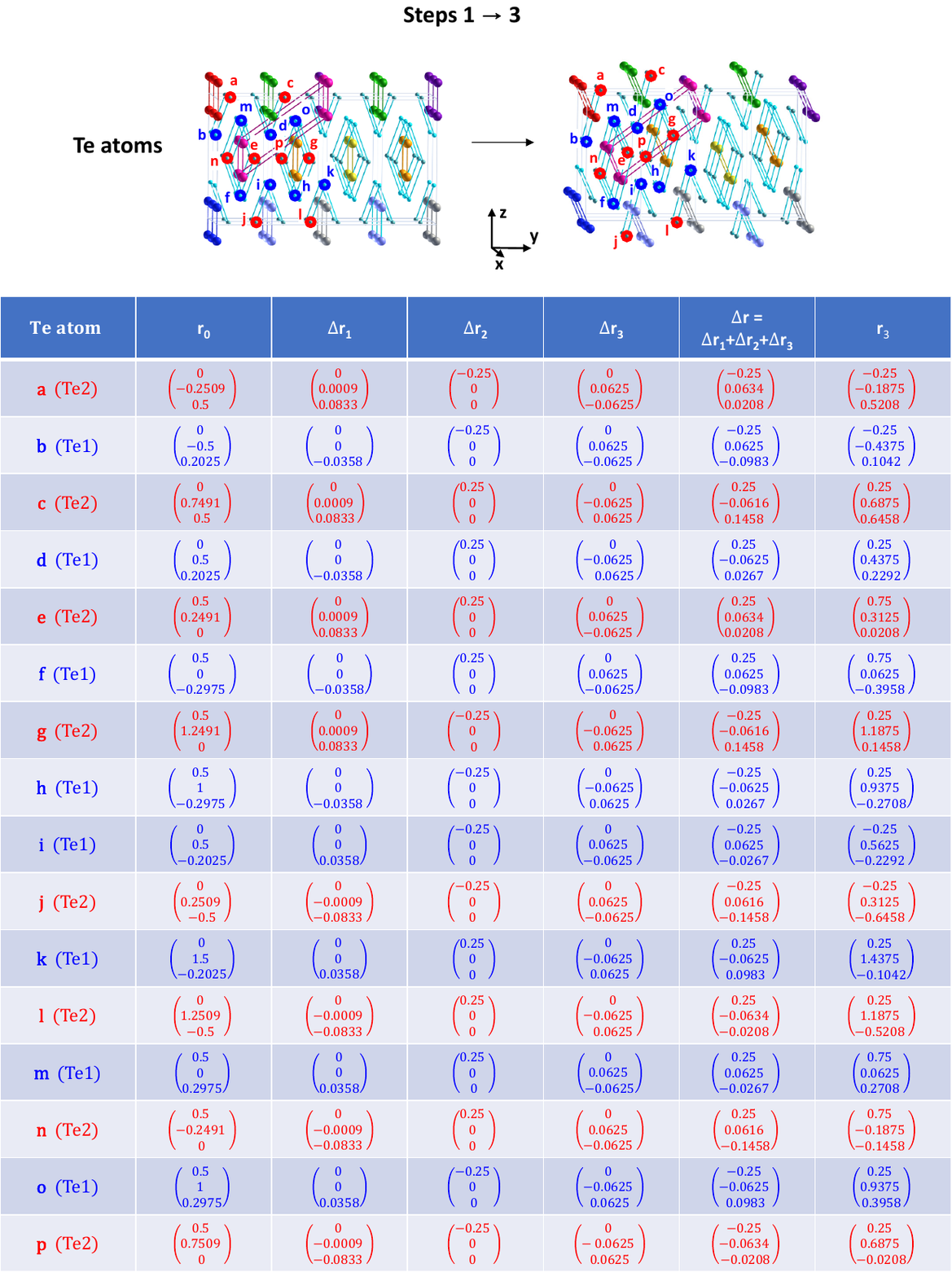}
\caption{\label{FigS17} (top) Identification of Te atoms noted \textbf{a} to \textbf{p} and (bottom) Table indicating their displacements in the steps 1 to 3, assuming the first scenario from [\onlinecite{Honda2023}] for the structural transition.}
\end{figure*}

We give here Supplementary information, in addition to that given in Section IVB of the main article, about the atomic displacements and lattice distortions at the structural transition induced in UTe$_2$ under pressure.  Different sets of colors are used in the Figures \ref{FigS13}, \ref{FigS14}, \ref{FigS15}, \ref{FigS16}, \ref{FigS17} to help visualizing the displacements of U atoms from different ladders.

Figure \ref{FigS13} shows three-dimensional views of the U-atoms lattice extended to several unit cells and illustrating the three scenarios proposed by Honda \textit{et al} to describe the structural transition of UTe$_2$ under pressure [\onlinecite{Honda2023S}]. In each scenario a given group of atoms in the orthorhombic structure, colored in magenta in Figure \ref{FigS13}, is identified as leading to an elementary cell of the tetragonal structure. Each scenario implies a different set of atom displacements and lattice distortions at the transition. The first scenario implies the smallest translations of the atoms and the smallest tilts of the atomic bonds at the structural transition, which contrasts with the more complex sets of atomic rearrangements needed for the second and third scenarios. In the first scenario, Te atoms already almost lie within planes $\perp\mathbf{n}$, which are transformed into the basal planes $\perp\mathbf{c}'$ of the tetragonal structure (see Figure 10 in the main paper). For these reasons, the first scenario proposed by Honda \textit{et al} seems more likely than the two others. In the following, we describe the structural transition using this scenario.

Figures \ref{FigS14} and \ref{FigS15} show that the structural transition can be artificially decomposed into five elementary steps corresponding to a given set of atomic displacements or lattice distortions. Figures \ref{FigS16} and \ref{FigS17} synthesize translations of U, Te1 and Te2 atoms within the Steps 1 to 3. 8 groups of equivalent U atoms and 16 groups of equivalent Te atoms are identified. All atoms from the same chain along $\mathbf{a}$ are equivalent and associated with a unique set of translations within the Steps 1 to 3. We can group these chains into elementary $\mathbf{a}$-axis patterns constituted of a two-leg ladder of U atoms and its four nearest chains of Te atoms. Within the Steps 2 and 3, each of these patterns are translated by a given vector in the plane perpendicular to $\mathbf{a}$.
The five steps are summarized below:
\begin{itemize}
  \item Step 1: U atoms forming the legs of the ladders are translated by $\mathbf{\Delta r}_1=\pm(0,0.125,0.01044)$, which corresponds to a tilt of the ladders ending in the arrangement of U atoms into planes of Miller indices (0 1 1). This translation ends into chains of equidistant U atoms along the direction (0,1,-1). Te1 atoms are translated by $\mathbf{\Delta r}_1=\pm(0,0,0.0358)$ and Te2 atoms are translated by $\mathbf{\Delta r}_1=\pm(0,0.0009,0.0833)$, ending also in the arrangement of Te atoms into planes of Miller indices (0 1 1) and chains of equidistant Te atoms along the direction (0,1,-1).
  \item Step 2 corresponds to a translation of the elementary  $\mathbf{a}$-axis patterns by $\mathbf{\Delta r}_2=\pm(0.25,0,0)$. It ends in square lattices of U atoms and Te atoms in the planes of Miller indices (0 1 1).
  \item Step 3 corresponds to a translation of the elementary $\mathbf{a}$-axis patterns by $\mathbf{\Delta r}_3=\pm(0,0.0625,-0.0625)$. It ends into a body-centered structure of the atoms.
  \item Step 4 corresponds to the modifications of the lattice parameters of the orthorhombic unit cell $a=4.16~\AA~\mapsto3.89$~\AA, $b=6.12~\AA~\mapsto7.78$~\AA, $c=13.96~\AA~\mapsto13.48$~\AA. It ends in a tetragonal structure with basal planes perpendicular to the direction $\mathbf{c}'$ and the lattice parameters $a'=3.89$~\AA and $c'=13.48$~\AA. These basal planes result from the transformation by Steps 1 to 4 of the planes of Miller indices (0 1 1) perpendicular to $\mathbf{n}$ defined for the orthorhombic structure.
  \item Step 5 corresponds to a compression along the direction $\mathbf{c}'$, ending in the modification of the lattice parameter $c'=13.48~\AA~\mapsto9.80$~\AA of the tetragonal structure.
\end{itemize}

\section{Supplementary discussion}
\label{comparison_SM}

In the electrical-resistivity studies presented here and in [\onlinecite{Valiska2021S},\onlinecite{Valiska2021SIS}] (electrical resistivity), signatures of the tetragonal phase were observed under pressures $p>3.1$~GPa, indicating a critical pressure $p^*\lesssim3$~GPa, for $\mathbf{H}$ tilted by $\simeq25-30~^\circ$ from $\mathbf{b}$ to $\mathbf{c}$. Both studies have been performed combining pulsed magnetic fields up to 60~T and pressures up to 6~GPa using the same pressure cell [\onlinecite{Braithwaite2016S},\onlinecite{Settai2015S}]. In [\onlinecite{Valiska2021S},\onlinecite{Valiska2021SIS}], the signatures of the high-pressure tetragonal phase, i.e., the transitions at $T_x$ and $H_x$, were already observed at pressures $p\geq3.1$~GPa in a magnetic field $\mathbf{H}\simeq\parallel\mathbf{b}$. However, a 15-30~$^\circ$ misorientation of the crystal relatively to the field direction was identified from a metamagnetic field $H_m$ much larger than that expected for $\mathbf{H}\parallel\mathbf{b}$. The raw electrical-resistivity data, the pressure variation of $H_m$, but also the field variations of $A$ found for $\mathbf{H}\simeq\parallel\mathbf{b}$ with 15-30~$^\circ$ misorientation in [\onlinecite{Valiska2021S},\onlinecite{Valiska2021SIS}] are strikingly similar to those obtained here, where the magnetic-field direction was carefully tilted by 30~$^\circ$ from $\mathbf{b}$ to $\mathbf{c}$. Retrospectively, we now suspect that, in the prior work published in [\onlinecite{Valiska2021S},\onlinecite{Valiska2021SIS}], the sample became aligned with a field $\mathbf{H}\simeq\parallel\mathbf{n}$, i.e., tilted by $\simeq25-30^\circ$ from $\mathbf{b}$ to $\mathbf{c}$ during the pressurization. Indeed the sample was set up in the cell with the plane $\perp\mathbf{c}$ placed vertically, but the upper and lower surfaces were cleaving surfaces $\perp\mathbf{n}$ which may well have become aligned with the anvil on pressurization. In comparison with the two sets of data collected in [\onlinecite{Valiska2021S},\onlinecite{Valiska2021SIS}], new information is extracted here concerning the high-pressure tetragonal phase. Stronger field-induced anomalies are observed at the lowest temperature ($T=1.4$~K), presumably due to an increase of the sample quality (the sample studied here was grown by the molten-salt-flux method [\onlinecite{Sakai2022}], while the samples studied in [\onlinecite{Valiska2021S},\onlinecite{Valiska2021SIS}]  were grown by the chemical-vapor-transport method [\onlinecite{Aoki2019bS}]).

\newpage

\end{document}